\documentclass[preprint,aoas]{imsart}

\RequirePackage[colorlinks,citecolor=blue,urlcolor=blue]{hyperref}
\usepackage{mathrsfs} 
\usepackage{amsmath} 
\usepackage{xcolor} 
\usepackage{multirow,booktabs,graphicx} 
\usepackage{pbox} 
\RequirePackage{natbib}

\startlocaldefs
\def\bX{\mathbf{X}}
\def\bx{\mathbf{x}}
\def\bXvec{\overrightarrow{\mathbf{X}}}
\def\bXbar{\bar{\mathbf{X}}}
\def\bI{\mathbf{I}}
\def\bJ{\mathbf{J}}
\def\bU{\mathbf{U}}
\def\bt{\mathbf{t}}
\def\bS{\mathbf{S}}

\def\Vhat{\widehat{V}}
\def\Ntld{\widetilde{N}}
\def\Nstr{N^*}

\def\balph{\boldsymbol\alpha}
\def\balphstr{\boldsymbol\alpha^*}
\def\balphstrtrans{\boldsymbol\alpha^{* \sf \tiny T}}
\def\balphtld{\widetilde{\boldsymbol\alpha}}
\def\balphhat{\widehat{\boldsymbol\alpha}}
\def\balphdag{\boldsymbol\alpha^{\dagger}}
\def\bbeta{\boldsymbol\beta}
\def\bmu{\boldsymbol\mu}
\def\bmustr{\boldsymbol\mu^*}
\def\bmutld{\widetilde{\bmu}}
\def\bmuhat{\widehat{\bmu}}
\def\mustr{\mu^*}
\def\mutld{\widetilde{\mu}}
\def\muhat{\widehat{\mu}}
\def\balphhat{\widehat{\balph}}
\def\Delthat{\widehat{\Delta}}

\def\Deltstr{\Delta^*}

\def\bgamma{\boldsymbol\gamma}
\def\bgammahat{\widehat{\boldsymbol\gamma}}
\def\bC{\boldsymbol{C}}
\def\bCtld{\widetilde{\boldsymbol{C}}}

\def\sigmahat{\widehat{\sigma}}
\def\Varhat{\widehat{Var}}

\def\varphitld{\widetilde{\varphi}}
\def\bvarphi{\boldsymbol\varphi}
\def\psitld{\widetilde{\psi}}

\def\Dscr{\mathscr{D}}
\def\Xscr{\mathcal{X}}

\def\Ybar{\bar{Y}}
\def\Xbar{\bar{\bX}}

\def\ddu{\frac{\partial}{\partial u}}
\def\ddbalph1{\frac{\partial}{\partial\balphtld_1}}
\def\ddbmu2{\frac{\partial}{\partial\bmutld_{\bX_2}}}
\def\ddbalphh{\frac{\partial^2}{\partial\balph_1^{\otimes 2}}}
\def\ddbmuu{\frac{\partial^2}{\partial\bmu_{\bX_2}^{\otimes 2}}}
\def\ddbalphmu{\frac{\partial^2}{\partial\balph_1\partial\bmu_{\bX_2}}}
\def\ddbalphT{\frac{\partial}{\partial\balph_1^{\trans}}}
\def\ddbmuT{\frac{\partial}{\partial\bmu_{\bX_2}\trans}}
\def\ddmu1{\frac{\partial}{\partial\mu_{1}}}

\def\indep{\perp\!\!\!\perp}
\def\trans{^{\sf \tiny T}}
\def\transdag{^{\dagger\sf \tiny T}}
\newcommand{\norm}[1]{\left\lVert#1\right\rVert}

\def\bzero{\mathbf{0}}
\def\bone{\mathbf{1}}

\definecolor{darkred}{RGB}{150,50,50}

\newenvironment{eq} 
{
\begin{equation}
\begin{aligned} 
}
{
\end{aligned}
\end{equation}
}

\newenvironment{eq*} 
{
\begin{equation*}
\begin{aligned} 
}
{
\end{aligned}
\end{equation*}
}

\newtheorem{assump}{Assumption}[section]
\newtheorem{theorem}{Theorem}[section]
\newtheorem{lemma}{Lemma}[section]
\endlocaldefs

\begin{document}

\begin{frontmatter}

\title{The Statistical Performance of Matching-Adjusted Indirect Comparisons}
\runtitle{Statistical Performance of MAIC}

\author{\fnms{David} \snm{Cheng}\thanksref{t1}},
\author{\fnms{Rajeev} \snm{Ayyagari}\thanksref{t2}}
\and
\author{\fnms{James} \snm{Signorovitch}\thanksref{t2}}

\affiliation{VA Boston Healthcare System\thanksmark{t1} and Analysis Group\thanksmark{t2}}

\begin{abstract}
Indirect comparisons of treatment-specific outcomes across separate studies often inform decision-making in the absence of head-to-head randomized comparisons.  Differences in baseline characteristics between study populations may introduce confounding bias in such comparisons.  Matching-adjusted indirect comparison (MAIC) \citep{signorovitch2010comparative} has been used to adjust for differences in observed baseline covariates when the individual patient-level data (IPD) are available for only one study and aggregate data (AGD) are available for the other study.  The approach weights outcomes from the IPD using estimates of trial selection odds that balance baseline covariates between the IPD and AGD.  With the increasing use of MAIC, there is a need for formal assessments of its statistical properties.  In this paper we formulate identification assumptions for causal estimands that justify MAIC estimators.  We then examine large sample properties and evaluate strategies for estimating standard errors without the full IPD from both studies.  The finite-sample bias of MAIC and the performance of confidence intervals based on different standard error estimators are evaluated through simulations.  The method is illustrated through an example comparing placebo arm and natural history outcomes in Duchenne muscular dystrophy.
\end{abstract}


\end{frontmatter}

\section{Introduction}
Indirect comparisons refer to comparisons of treatment-specific outcomes across different studies, as opposed to direct comparisons of treatments-specific outcomes within a single study.  This frequently arises in clinical, economic, and regulatory evaluations of new treatments \citep{altman2005indirect,sutton2008use,wells2009indirect,jansen2011interpreting}. Even when a new treatment has been compared with a standard-of-care in a randomized trial, for example, it is often important for decision-makers to contextualize expected outcomes with other treatments and external data sources.
The need for indirect comparisons is also especially acute in settings where direct randomized comparisons are unethical or infeasible, such as in late-stage oncology trials \citep{adjei2009novel}, rare disease trials, and evaluations of long-term outcomes in open-label extension trials. 
Indirect comparisons also inform the design of trials by providing reasonable prior estimates of effect sizes for power calculations and non-inferiority margins \citep{snapinn2011indirect}.

Since indirect comparisons involve comparisons between studies for which assignment to different studies is not randomized, 
the resulting treatment effect estimates may be confounded by cross-study differences in baseline characteristics. 
Such differences may exist even when treatment groups \emph{within} each study have been randomized. In practice, a host 
of differences in the design and setting of each study can bias indirect comparisons. It is essential to consider, with clinical 
input, definitions and assessment methodologies for the outcome measure, patient selection criteria, 
background care, along with other issues
\citep{pocock1976combination,us2001guidance,phillippo20162016}.   A non-exhaustive list of considerations is provided in Table \ref{t:desgnbias}.
 Although 
a wide range of issues may introduce bias, there are many cases in which separate studies 
are found, after careful evaluation, to be sufficiently similar 
for an indirect comparison. Registrational trials conducted for different treatments within the same indication during similar time periods, for instance, often share a high degree of similarity.
In this paper we assume that trials are sufficiently similar in design such that attention can be focused on addressing bias that stems from differences in baseline characteristics.

\begin{table}[htbp]
  \centering
  \caption{Study design features that may vary across studies and bias indirect comparisons.  Differences in observed baseline characteristics, the final category, are addressed by the adjustment methods described in this paper.}
    \begin{tabular}{l}
    \toprule
    \textbf{Aspect of study design} \\
    \midrule
    \textit{Outcome assessment} \\
    \multicolumn{1}{l}{\qquad Clinical, imaging, or laboratory methods used to process and measure outcomes} \\
    \multicolumn{1}{l}{\qquad Outcome definitions, including timing of assessments} \\
    \multicolumn{1}{l}{\qquad Event ascertainment or adjudication procedures} \\
    \multicolumn{1}{l}{\qquad Completeness of follow-up, reasons for drop-out} \\
    \textit{Patient selection} \\
    \multicolumn{1}{l}{\qquad Inclusion and exclusion criteria} \\
    \multicolumn{1}{l}{\qquad Recruitment process (e.g. motivation, consent process, distance of site to patient)} \\
    \multicolumn{1}{l}{\qquad Disease diagnostic criteria} \\
    \textit{Background treatments and care settings} \\
    \multicolumn{1}{l}{\qquad Time period and geography, and associated standards of care} \\
    \multicolumn{1}{l}{\qquad Range of non-study treatment options available} \\
    \multicolumn{1}{l}{\qquad Concomitant medications} \\
    \textit{Baseline patient characteristics} \\
    \multicolumn{1}{l}{\qquad Demographics} \\
    \multicolumn{1}{l}{\qquad Comorbidities} \\
    \multicolumn{1}{l}{\qquad Treatment history} \\
    \multicolumn{1}{l}{\qquad Severity and duration of medical condition} \\
    \multicolumn{1}{l}{\qquad Biomarkers} \\
    \bottomrule
    \end{tabular}%
  \label{t:desgnbias}%
\end{table}%

When individual patient-level data (IPD) are available for both studies, the pooled data can essentially be regarded as 
observational data with a non-randomized treatment assignment.  In this case methods for estimating treatment effects 
with non-randomized treatment based on outcome regression and propensity score approaches are well-established \citep{lunceford2004stratification, kang2007demystifying}.  
Conducting indirect comparisons with full IPD can also be viewed as generalizing estimates from one study to a different 
target study population \citep{stuart2011use,hartman2015sample,nie2013likelihood,zhang2015new}, which has a long history 
of application in regulatory settings in the form of studies with historical controls \citep{us2001guidance,committee2006guideline}.  
Access to full IPD should always be the preferred approach to comparative analyses, when possible.

In practice, however, many indirect comparisons are conducted in settings where the full IPD are not available for both study populations.  When only aggregate data (AGD) consisting of summary statistics (e.g. means and standard errors from publications) for outcomes and baseline covariates are available, \cite{bucher1997results} introduced a widely used method to compare treatment effects on an appropriate scale of contrast (e.g. log odds-ratio, risk difference, etc.) relative to a common comparator group.  But, as we discuss in Section \ref{s:sim}, this method requires a common comparator arm between studies and 
does not eliminate the risk of bias.
If AGD from a large number of studies are available, mixed treatment comparisons generalize methods from meta-analysis and \cite{bucher1997results} to allow for comparisons of treatments within a network of randomized trials that are linked by common comparators \citep{lu2004combination,nixon2007using}.
These approaches rely on ``consistency'' or ``exchangeability'' assumptions, requiring treatment effects to be constant across trials on a specified scale of contrast.  These assumptions are typically violated when study populations differ in baseline characteristics, particularly those that modify treatment effects on the chosen contrast scale.

It is often the case that IPD are available for one study whereas only AGD are available for others.  This can occur, 
for example, when researchers can access IPD from a study they conducted but cannot directly access IPD underlying published 
AGD from other studies.
When IPD are available for only one study and AGD are available for another, two general approaches have been used. The simulated treatment comparison (STC) method estimates an adjusted mean outcome under the IPD treatment by fitting an outcome regression model to the IPD and plugging-in baseline covariates from the AGD \citep{caro2010no,ishak2015simulation}. Indirect comparisons can be obtained by contrasting the predicted value with observed mean outcomes in the AGD.
If a non-linear regression model is used, such as a logit link in a logistic regression model or a log link in a proportional hazards model, this method incurs bias because expectations do not commute with non-linear functions. Addressing this requires parametric assumptions about the full joint distribution of covariates from the AGD.
STC could also be biased when the postulated outcome regression model is mis-specified.  An additional method is matching-adjusted indirect comparison (MAIC) \citep{signorovitch2010comparative}, which estimates mean outcomes for the IPD in the population represented by the AGD by estimating trial selection odds through a method of moments approach and re-weighting the IPD. This avoids issues that arise from regression modeling, though it still generally assumes correctly specified models for the trial selection odds.
MAIC is similar in spirit to a number of other re-weighting methods that seek to balance covariates between treatment groups to estimate causal effects \citep{zubizarreta2015stable,li2017balancing} and identical to the covariate balancing propensity score used to estimate average treatment effects on the treated \citep{imai2014covariate} and entropy-balancing \citep{hainmueller2012entropy,zhao2017entropy} when a logistic regression model is assumed for the trial assignment model (see Section \ref{ss:ident}).
However, these other methods are largely motivated by concerns about the balancing performance of propensity scores estimated from parametric models and focus on estimation and inference in other settings where IPD are fully observed.

Although MAIC has been successfully applied in health technology assessments, in published outcomes research studies \citep{signorovitch2012matching,signorovitch2013everolimus,phillippo20162016,swallow2016daclatasvir}, 
and in clinical regulatory evaluations \citep{ema2018kymriah}, few formal evaluations of its underlying assumptions and statistical properties 
exist.  We aim to partly address this gap.  Specifically, we formalize the problem in the framework of counterfactual 
outcomes and formulate identification assumptions for causal estimands in the setting where IPD is available for only one 
study in Sections \ref{ss:setup} and \ref{ss:ident}.  We then study the large sample properties of the MAIC estimator in 
Section \ref{ss:estimation} and discuss strategies to estimate the standard error in the absence of IPD from the AGD trial 
in Section \ref{ss:estvar}.  An investigation of the finite sample performance of the MAIC estimator and standard error 
estimators is reported in Section \ref{s:sim}.  We illustrate the method through an application to Duchenne muscular 
dystrophy in Section \ref{s:application} and conclude with some further discussion in Section \ref{s:discuss}.  
Proofs are deferred to Appendix \ref{apd:proofs}.

\section{Method}
\subsection{Problem Setup and Notation}
\label{ss:setup}
For each individual $i$ in either the IPD or AGD study, let $Y_i$ denote a continuous or binary outcome and $\bX_i$ a $p$-dimensional vector of baseline covariates belonging to covariate space $\Xscr$.  Let $Z_i \in \{ 0,1,2\}$ denote treatment assignment to either a common comparator $(Z_i=0)$ or a treatment studied only in the IPD or AGD trial ($Z_i=1$ or $Z_i=2$, respectively), which is randomized within each trial.  Let $T_i \in \{ 1,2\}$ denote trial assignment to the IPD trial ($T_i=1$) or AGD trial ($T_i = 2$).  A common comparator is assumed to be available here to facilitate some parts of the subsequent exposition, but it is not strictly needed for MAIC.  In cases with single arm studies, the common comparator data is omitted and $Z_i = T_i$.  The IPD, in general, thus consists of independent and identically distributed (iid) observations $\Dscr_{IPD} = \left\{ (Y_i, Z_i,\bX_i) : T_i = 1\right\}$.  The AGD consists of data summaries $\Dscr_{AGD}=\left\{ \Ybar_{2z}, \Xbar_{2z}, S^2_{Y,2z}, \bS^2_{\bX,2z}, N_{2z}: z = 0,2\right\}$, where $N_{tz} = \sum_{T_i=t}I(Z_i=z)$, $\Xbar_{tz} = \sum_{T_i=t}\bX_iI(Z_i=z) /N_{tz}$, and $\Ybar_{tz} = \sum_{T_i=t}Y_iI(Z_i=z) /N_{tz}$.  Among arm $z$ of trial $t$, the sample variance of the outcome is $S^2_{Y,tz} = \sum_{T_i=t}(Y_i-\Ybar_{tz})^2I(Z_i=z)/(N_{tz}-1)$ and of all covariates is $\bS^2_{\bX,tz} = \sum_{T_i=t}(\bX_i - \Xbar_{tz})^{\odot 2}I(Z_i=z)/(N_{tz}-1)$, with $\odot$ denoting element-wise exponentiation.  The sample covariance matrix for $\bX$ is not fully available since the covariance between covariates is typically not reported in publications.  Let the total size between trials be $N=\sum_{t,z}N_{tz}$.  

The goal of an indirect comparison is to conduct a ``fair'' comparison of the mean outcomes under treatment $1$ to treatment $2$, 
ideally accounting for differences in outcomes due to discrepancies in the distribution of $\bX$ between studies.
When IPD on both baseline covariates $\bX$ and outcomes $Y$ are available for a group of patients, it is potentially possible to
re-estimate their mean outcomes while adjusting the distribution of $\bX$ to more closely match that of another 
population with sufficiently overlapping support.  As IPD is available for patients 
treated with treatment $1$, we can adjust their mean outcomes to more closely match the distribution of $\bX$ of those who received treatment
$2$, i.e. the $T=2$ population, but not vice versa, since only AGD is available for patients who received treatment $2$.  
Let $Y_i(z)$ denote the counterfactual outcome had patient $i$ been treated with treatment $z$.  
With these considerations in mind, we take the target estimand to be:
\begin{eq} \label{e:targ}
\Delta = E\left\{Y(1) | T=2\right\} - E\left\{Y(2) | T=2\right\}.
\end{eq}
This is the average treatment effect on the treated (ATT) among those assigned to treatment $2$.  
When a different scale for the treatment effect contrast is of interest, we can consider a generalization:
\begin{eq*}
\Delta_g = g\left(E\left\{Y(1) | T=2\right\}\right) - g\left(E\left\{Y(2) | T=2\right\}\right),
\end{eq*}
where $g(\cdot)$ is a given link function.  For example, when $Y$ is binary, taking $g(u) = log\{u/(1-u)\}$ specifies that $\Delta_g$ is on the log odds-ratio scale (LOR).  Regardless of the scale of contrast, the main challenge will be to identify and estimate $E\left\{Y(1) | T=2\right\}$.  We will focus on $\Delta$ as the target parameter for conciseness and note that a transformation can be applied to obtain $\Delta_g$.

\subsection{Identification}
\label{ss:ident}
As $\Delta$ is defined in terms of unobserved counterfactual outcomes, we consider the following assumptions required for identification.
\begin{assump}
\label{a:randm}
Random treatment within trial:
$Z \indep \{\bX, Y(1), Y(0)\} |T=1$ and $Z \indep \{\bX, Y(2),Y(0)\} |T=2$.
\end{assump}
\begin{assump}
\label{a:consi}
Consistency: $Y = Y(Z)$ with probability $1$.
\end{assump}
\begin{assump}
\label{a:posit}
Positivity of trial assignment: $P(T=1|\bX=\bx) \geq \epsilon_{T|\bX}$ for all $\bx \in \Xscr$, for some $\epsilon_{T|\bX} >0$.
\end{assump}
\begin{assump}
\label{a:ignor}
Ignorability of trial assignment for the outcome under treatment $1$:
$T \indep Y(1) |\bX$.
\end{assump}
The first assumption refers to the independence of treatment assignment with covariates and counterfactual outcomes within each trial due to randomization.  This assumption would be omitted in the case of single-arm studies.  The consistency assumption states that the observed outcome coincides with the counterfactual outcome under the treatment received, which excludes settings with interference and different versions of a treatment.  The positivity assumption indicates patients are not assigned to the AGD trial using a deterministic or nearly deterministic rule in $\bX$.  
This is a non-trivial assumption that could be violated, for instance, when inclusion/exclusion criteria are such that the 
AGD trial includes patients who are excluded from the IPD trial.  In such cases, the underlying trial populations are not 
similar enough to conduct an indirect comparison with any method, without extrapolating beyond the population of the available 
IPD.  Conversely, the assumption allows for $P(T=2|\bX=\bx) = 0$ for some $\bx \in \Xscr$ so that there may be patients
enrolled in the IPD trial who would be excluded from the AGD trial.  In theory such patients can simply be excluded to achieve
balance in the distribution of $\bX$ between studies.  We offer some more discussion on the effects of violating this 
assumption in practice in Section \ref{s:discuss}. 
The trial assignment ignorability assumption states that we observe sufficient covariates $\bX$ such that trial assignment 
is unrelated to the counterfactual outcome under treatment $1$ within strata of $\bX$.  This is plausible when the patient 
populations between the two trials are similar enough such that conditioning on the observed covariates $\bX$ is enough 
to control for the differences between study populations that could lead to confounding bias.  As adjustments for the
the distribution of $\bX$ is applied to only the IPD estimates, ignorability is required only for treatment $1$ and not
for treatment $2$.

Based on these assumptions, the second term of $\Delta$ can be identified as $E\left\{Y(2)|T=2\right\}=E(Y|Z=2,T=2)$ using 
Assumptions \ref{a:randm} and \ref{a:consi}.  The first term of $\Delta$ can be identified as:
\begin{eq}
\label{e:id}
E&\left\{Y(1)|T=2\right\} 
= E\left[E\left\{Y(1)|\bX,T=2\right\}|T=2\right] \\
&= E\left[E\left\{Y|\bX,Z=1,T=1\right\}|T=2\right] \\
&= E\left\{\frac{P(T=1)}{P(T=2)}\frac{P(T=2|\bX)}{P(T=1|\bX)}E(Y|T=1,\bX,Z=1)|T=1\right\} \\
&= E\left\{\frac{I(Z=1)}{P(Z=1|T=1)}\frac{I(T=1)}{P(T=2)}\omega(\bX)Y\right\} \\
&= E\left\{I(Z=1,T=1)\omega(\bX)Y\right\}/E\left\{ I(Z=1,T=1)\omega(\bX)\right\},
\end{eq}
where $\omega(\bX) = P(T=2|\bX)/P(T=1|\bX)$ denotes the odds of trial assignment given $\bX$.  We used Assumptions \ref{a:randm}, \ref{a:consi}, \ref{a:ignor} in the second equality, \ref{a:posit} in the third and fourth equalities, and \ref{a:randm} in the fourth and final equalities.  This suggests that $\Delta$ can be identified if $\omega(\bX)$ can be identified.  Even if $\omega(\bX)$ could not be identified exactly, the last equality shows it would be sufficient if $\omega(\bX)$ could be identified up to a constant due to the ratio of terms involving $\omega(\bX)$.

Before proceeding, we first discuss an alternative to Assumption \ref{a:ignor} that leverages the common comparator arm when data on it is available.
\begin{assump}
\label{a:ignormod}
Ignorability of trial assignment for the counterfactual difference:
$T \indep \left\{Y(1)-Y(0)\right\} |\bX$.
\end{assump}
To make use of this assumption, the target parameter can be written $\Delta = E\left\{ Y(1) - Y(0) |T=2\right\} - E\left\{ Y(2) - Y(0) |T=2\right\}$, where $E\left\{ Y(2) - Y(0) |T=2\right\}$ is straightforward to identify due to randomization of treatment within trial.  The first term can be identified by:
\begin{eq}
\label{e:idalt}
E&\left\{ Y(1) - Y(0) |T=2\right\} = E\left[E\left\{ Y(1) - Y(0) |\bX, T=2\right\}|T=2\right] \\
&= E\left[E\left\{ Y(1) - Y(0) |\bX, T=1\right\}|T=2\right] \\
&= E\left[E\left\{ Y |\bX, Z=1,T=1\right\} - E\left\{Y|\bX, Z=0, T=1\right\} |T=2\right] \\
&= E\left\{\frac{I(Z=1)}{P(Z=1|T=1)}\frac{I(T=1)}{P(T=2)}\frac{P(T=2|\bX)}{P(T=1|\bX)}Y\right\} \\
&\qquad - E\left\{\frac{I(Z=0)}{P(Z=0|T=1)}\frac{I(T=1)}{P(T=2)}\frac{P(T=2|\bX)}{P(T=1|\bX)}Y\right\}
\end{eq}
where we use Assumption \ref{a:ignormod} in the second equality, \ref{a:randm} and \ref{a:consi} in the second equality.  The final steps proceeds as in \eqref{e:id}, which uses Assumptions \ref{a:posit} and \ref{a:randm}.  This alternative assumption could be  easier to justify in some cases.  For example, suppose the outcome model can be specified by:
\begin{eq*}
E(Y|Z,\bX)=\beta_0 + \beta_1 I(Z\neq 0) + \bbeta_2\trans \bX + \bbeta_3\trans I(Z\neq 0)\bX_M + \beta_4I(Z=2),
\end{eq*}
where $\bX_M \subseteq \bX$ is a subset of the prognostic covariates that modify the treatment effect of the active treatments and the components of $\bbeta_3$ are not exactly the negative of the components of $\bbeta_2$ corresponding to the same covariate in $\bX_M$. In such cases it suffices to adjust only for 
the
treatment effect modifiers $\bX_M$ rather than the full set of $\bX$ that may differ in distribution between trials \citep{phillippo20162016}.  This approach can also be used to identify $\Delta_g$ for a non-linear $g(\cdot)$ by writing $E\left\{ Y(1)|T=2\right\} = E\left\{ Y(1) - Y(0) |T=2\right\} + E\left\{Y(0) |T=2\right\}$.  The second term can be easily identified from randomization within trial. This approach offers an alternative identification strategy but still relies on strong assumptions about the form of the conditional mean of $Y$ and assumes covariates are known not to be effect modifiers on the additive scale.
Despite its appeal, Assumption \ref{a:ignormod} is not necessarily weaker than \ref{a:ignor} and should be evaluated based on clinical input to the extent possible in practice.

It now remains for us to identify $\omega(\bX)$ from the observed data.  Had IPD been available for both trials, then this would be straightforward.  In the absence of the full IPD, under Assumption \ref{a:posit}, $\omega(\bX)$ is a solution in $h(\bX)$ to the integral equation:
\begin{eq}
\label{e:momcnd}
E\left\{ h(\bX) I(T=1)\bXvec\right\}-\bmustr_{\bXvec_2} P(T=2) = \bzero,
\end{eq}
where $\bXvec = (1,\bX\trans)\trans$ and $\bmustr_{\bXvec_2}=E(\bXvec|T=2)$.  If $h^*(\bX)$ is a solution to this equation and the solution is \emph{unique}, then it must be that $h^*(\bX)=\omega(\bX)$.  In general, without any restrictions on $h(\bX)$, there may not be a unique  solution.
However, $h(\bX)$ can be reasonably parameterized such that there exists a unique solution.  For example, suppose the trial assignment follows a logistic regression model:
\begin{eq}
\label{e:trlmod}
logit P(T=2|\bX) = \balph\trans\bXvec,
\end{eq}
for some $\balph = (\alpha_0,\balph_1\trans)\trans$.  Then under this model $\omega(\bX) = exp( \balphstrtrans\bXvec)$, where $\balphstr$ is the true value of $\balph$.  Restricting $h(\bX) = \omega(\bX;\balph) = exp(\balph\trans\bXvec)$,  \eqref{e:momcnd} admits a unique solution because $Q(\balph)=E\left\{ \omega(\bX;\balph) I(T=1)\right\} - \balph\trans\bmustr_{\bXvec_2}P(T=2)$ is strictly convex in $\balph$ \citep{signorovitch2010comparative}.  

Other moment conditions can also be used.  Since $\omega(\bX)$ needs only to be identified up to a constant, an alternative condition is:
\begin{eq}
\label{e:momcndalt}
E\left\{ h(\bX-\bmustr_{\bX_2}) I(T=1)(\bX-\bmustr_{\bX_2})\right\} = \bzero,
\end{eq}
where $\bmustr_{\bX_2} = E(\bX|T=2)$. Under model \eqref{e:trlmod}, $\balph_1$ can be identified by similar arguments by restricting $h(\bX-\bmu_{\bX_2})=\omega(\bX-\bmu_{\bX_2};\balph_1)=exp\{ \balph_1\trans(\bX-\bmu_{\bX_2})\}$.  { This approach identifies $\omega(\bX)$ up to a scalar constant and avoids estimation of an additional intercept parameter that is not needed for estimating $\Delta$.}  Another potentially useful condition is:
\begin{eq}
\label{e:momcndvar}
E\left\{ h(\bX)I(T=1) \bt(\bX)\right\} - E\left\{\bt(\bX)|T=2\right\}P(T=2)=\bzero,
\end{eq}
where we take $\bt(\bX) = (1, X_{1},\ldots,X_p,X_1^2,\ldots,X_p^2)\trans$ to be elements of $\bX$ and their squares.  In contrast to conditions \eqref{e:momcnd} and \eqref{e:momcndalt}, using this condition in practice would require the availability of the sample variances of the covariates in the AGD to estimate $E\{\bt(\bX)|T=2\}$.  If we again restrict $h(\bX) = \omega(\bX;\balphdag) = exp\{ \balph\transdag\bt(\bX)\}$, where $\balphdag = (\balph\trans,\balph_2\trans)\trans$, then $\omega(\bX)$ is again identified under the more general trial assignment model:
\begin{eq*}
logitP(T=2|\bX) = \balph\transdag\bt(\bX).
\end{eq*}
The model from \eqref{e:trlmod} can be viewed as a submodel of this model that restricts $\balph_2=\bzero$.  When \eqref{e:trlmod} is correct, fitting this larger model comes at the cost of finite-sample efficiency loss.
However, if this expanded model is correct, then using \eqref{e:momcnd} for estimation would yield biased estimates of $\omega(\bX)$ and $\Delta$.  Balancing the first and second moments with this expanded model thus involves trade-off between robustness and efficiency.  We next discuss estimation of $\omega(\bX)$ and $\Delta $ based on these identification conditions.

\subsection{Estimation}
\label{ss:estimation}
For the rest of the paper, except for Section \ref{ss:estvar}, we will assume a model where trial assignment is correctly specified by \eqref{e:trlmod}.  We first estimate $\omega(\bX)$ by solving empirical versions of the proposed moment conditions.  For instance, let $\balphhat_1$ be the solution to the equation:
\begin{eq}
\label{e:alphee}
N^{-1}\sum_{i=1}^N \omega(\bX_i-\bXbar_2;\balph_1)I(T_i=1)(\bX_i-\bXbar_2) = \bzero,
\end{eq}
where $\bXbar_2 = (\bXbar_{22}N_{22}+\bXbar_{20}N_{20})/(N_{22}+N_{20})$.  
{ We use the moment condition from \eqref{e:momcndalt} to avoid estimating the additional intercept parameter.}  The following result states that $\balphhat_1$ is consistent and asymptotically linear and thus also asymptotically normal.
\begin{theorem}
\label{thm:alphIF}
If the trial assignment model is correctly specified by \eqref{e:trlmod}, then $\balphhat_1  \overset{p}{\to} \balphstr_1$, where $\balphstr_1$ is the true coefficient in \eqref{e:trlmod}.  Furthermore, $\balphhat_1$ is asymptotically linear such that:
\begin{eq}
&N^{1/2}(\balphhat_1-\balphstr_1) = N^{-1/2} \sum_{i=1}^N \bvarphi_{i}^{\balph_1}(\balphstr_1,\bmustr_{\bX_2}) + o_p(1),
\end{eq}
where $\bvarphi^{\balph_1}_{i}(\balph_1,\bmu_{\bX_2}) = \bJ^{\balph_1}(\balph_1,\bmu_{\bX_2})^{-1}\left\{ \bU_i^{\balph_1}(\balph_1,\bmu_{\bX_2}) + \bU_i^{\bmu_{\bX_2}}(\bmu_{\bX_2},\balph_1)\right\}$ is a mean zero random vector with finite variance and:
\begin{eq*}
&\bU_i^{\balph_1}(\balph_1,\bmu_{\bX_2}) = (\bX_i - \bmu_{\bX_2})exp\{ \balph_1\trans(\bX_i - \bmu_{\bX_2})\}I(T_i = 1) \\
&\bU_i^{\bmu_{\bX_2}}(\bmu_{\bX_2},\balph_1) = - E[exp\{ \balph_1\trans(\bX_i-\bmu_{\bX_2})\}I(T_i=1)](\bX_i -\bmu_{\bX_2})\frac{I(T_i=2)}{P(T_i=2)} \\
&\bJ^{\balph_1}(\balph_1,\bmu_{\bX_2}) = -E[ (\bX_i - \bmu_{\bX_2})(\bX_i - \bmu_{\bX_2})\trans exp\{ \balphtld_1\trans(\bX_i-\bmu_{\bX_2})\}I(T_i=1)].
\end{eq*}
\end{theorem}
This expansion reveals two sources that contribute to the asymptotic variance.  Suppressing implicit arguments for the parameters, the $\bU_i^{\balph_1}$ term is contributed from estimating $\balph_1$ when $\bmu_{\bX_2}$ is known.  The $\bU_i^{\bmu_{\bX_2}}$ term is the additional contribution when $\bmu_{\bX_2}$ is considered to be estimated by $\bXbar_2$.  Though this expansion clarifies the sources of variability, the influence function cannot be directly used to compute the asymptotic variance since $\bU_i^{\bmu_{\bX_2}}$ involves IPD from the AGD trial.  

Following estimation of $\balph_1$, $\Delta$ can subsequently be estimated by an empirical version of \eqref{e:id}, plugging in $\omega(\bX;\balphhat_1)$ for $\omega(\bX)$.  We consider identification based on \eqref{e:id} instead of \eqref{e:idalt} for conciseness of presentation, but similar results will hold if \eqref{e:idalt} is used.  Specifically, the estimator can be expressed as:
\begin{eq}
\label{e:deltdef}
\Delthat &= \sum_{i=1}^N I(Z_i=1,T_i=1)\omega(\bX_i;\balphhat_1)Y_i/ \sum_{i=1}^N I(Z_i=1,T_i=1)\omega(\bX_i;\balphhat_1) \\
&\qquad\qquad - \Ybar_{22}.
\end{eq}
The following result states that $\Delthat$ is consistent and asymptotically linear and thus also asymptotically normal.
\begin{theorem}
\label{thm:DeltIF}
Suppose that the identification assumptions \eqref{a:randm}, \eqref{a:consi}, \eqref{a:posit}, and \eqref{a:ignor} hold and the trial assignment model is correctly specified by \eqref{e:trlmod}.  Then $\Delthat\overset{p}{\to}\Deltstr$, where $\Deltstr$ is the true target parameter $\Delta$.  Furthermore, $\Delthat$ is asymptotically linear such that:
\begin{eq}
\label{eq:fullIF}
N^{1/2}(\Delthat - \Deltstr) &= N^{-1/2}\sum_{i=1}^N \varphi_i(\Deltstr,\mustr_1,\balphstr_1,\bmustr_{\bX_2}) + o_p(1),
\end{eq}
where $\mustr_1 = E\{Y(1)|T=2\}$ is the true counterfactual mean for the IPD treatment in the AGD population
and $\varphi_i(\Delta,\mu_1,\balph_1,\bmu_{\bX_2}) = \varphi_i^{\mu_2}(\Delta,\mu_1) + \varphi_i^{\mu_1}(\mu_1,\balph_1) + J^{\mu_1}(\balph_1)^{-1}\bCtld_1(\mu_1,\balph_1)\trans\bvarphi_i^{\balph_1}(\balph_1,\bmu_{\bX_2})$ is a mean zero random variable with finite variance and:
\begin{eq*}
&\varphi_i^{\mu_2}(\Delta,\mu_1) = (\mu_1-Y_i -\Delta) \frac{I(Z_i = 2, T_i = 2)}{P(Z_i = 2, T_i = 2)} \\
&\varphi_i^{\mu_1}(\mu_1,\balph_1) = J^{\mu_1}(\balph_1)^{-1} (Y_i - \mu_1)e^{\balph_1\trans\bX_i}I(Z_i = 1,T_i=1)\\
&J^{\mu_1}(\balph_1) = E\{ e^{\balph_1\trans\bX_i}I(Z_i=1,T_i=1)\}\\
&\bCtld_1(\mu_1,\balph_1) = E\{\bX_i (Y_i -\mu_1)e^{\balph_1\trans\bX_i}I(Z_i=1,T_i=1)\}.
\end{eq*}
\end{theorem}
The first $\varphi_i^{\mu_2}$ term accounts for estimation of $\mustr_2=E\{Y(2)|T=2\}$ from $\Ybar_{22}$.  The subsequent terms account for estimating $\mustr_1$ through weighting, which can be further decomposed into a term contributed for estimating $\mustr_1$ when $\balph_1$ is known and another term for estimating $\balph_1$.  Again the asymptotic variance cannot be directly computed from this influence function because $\varphi_i^{\mu_2}$ and $\bvarphi_i^{\balph_1}$ involve the IPD from the AGD trial.  We argue in Section \ref{ss:estvar} that it is often sufficient to compensate for ignoring this term by simply incorporating the marginal variance of $\Ybar_{22}$ from the AGD trial and consider other potential strategies.

\subsection{Estimation of Asymptotic Variance} \label{ss:estvar}
Estimating the asymptotic variance of $\Delthat$ is complicated by the fact that $\Delthat$ depends on $\bXbar_2$ and $\Ybar_{22}$, 
and the IPD is not available to estimate contributions that account for their variability.  As we discuss in Appendix \ref{apd:altsamp}, though 
one can regard $\bmustr_{\bX_2}=\bXbar_2$ and $\mustr_2 = \Ybar_{22}$ to be fixed in the sampling scheme, this may not always
be justifiable.  In the following we consider strategies to estimate the full asymptotic variance, regarding $\bXbar_2$ and $\Ybar_{22}$
as random, in the absence of the full IPD. To facilitate the subsequent considerations, we first define the following contributions 
to the influence function for $\Delthat$  for estimating $\balph_1$ and $\bmu_{\bX_2}$:
\begin{eq*}
&\varphitld_i^{\balph_1}(\balph_1,\mu_1,\bmu_{\bX_2}) = J^{\mu_1}(\balph_1)^{-1} \bCtld_1(\mu_1,\balph_1)\trans \bJ^{\balph_1}(\balph_1,\bmu_{\bX_2})^{-1} \bU_i^{\balph_1}(\balph_1,\bmu_{\bX_2})\\
&\varphitld_i^{\bmu_{\bX_2}}(\bmu_{\bX_2},\mu_1,\balph_1) = J^{\mu_1}(\balph_1)^{-1}\bCtld_1(\mu_1,\balph_1)\trans \bJ^{\balph_1}(\balph_1,\bmu_{\bX_2})^{-1}\bU_i^{\bmu_{\bX_2}}(\balph_1,\bmu_{\bX_2}).
\end{eq*}
By calculating the variance of $\varphi_i(\Deltstr,\mustr_1,\balphstr_1,\bmustr_{\bX_2})$, the full asymptotic variance of $\Delthat$ can now be expressed as:
\begin{eq}
\label{e:sig2}
\sigma^2 &=  \left\{Var( \varphi_i^{\mu_1}) + Var( \varphi_i^{\mu_2}) \right\} + \left\{Var( \varphitld_i^{\balph_1}) + 2Cov( \varphi_i^{\mu_1}, \varphitld_i^{\balph_1})\right\} \\
&\qquad + \left\{Var( \varphitld_i^{\bmu_{\bX_2}}) + 2Cov( \varphi_i^{\mu_2}, \varphitld_i^{\bmu_{\bX_2}})\right\},
\end{eq}
where the arguments of the components of the influence function are suppressed but implicitly evaluated at their respective truth. Decomposing the variance this way, the first two terms constitutes the asymptotic variance had $\balph_1$ been known.  The second two terms are contributed from estimating $\balph_1$ had $\bmu_{\bX_2}$ been known, and final two terms account for estimating $\bmu_{\bX_2}$. 

It is generally not possible to fully estimate $\sigma^2$ without further assumptions as $\varphi_i^{\mu_2}$ and $\varphitld_i^{\bmu_{\bX_2}}$ involve IPD from the AGD trial. However, it may still be possible to obtain reasonable approximations. The following lemma further clarifies the form of the additional contributions from estimating $\balph_1$ and $\bmu_{\bX_2}$ under correct identification and modeling assumptions.
\begin{lemma}
\label{lem:varsmp}
Let identification assumptions \eqref{a:randm}, \eqref{a:consi}, \eqref{a:posit}, and \eqref{a:ignor} be satisfied and the trial assignment model be correctly specified by \eqref{e:trlmod}. The terms contributed from estimating $\balph_1$ from \eqref{e:sig2} can be simplified as:
\begin{eq*}
&Var( \varphitld_i^{\balph_1}) = P(T=2)^{-1}\bC_1\trans Var(\bX|T=2)^{-1} \\
&\quad E\{ (\bX - \bmustr_{\bX_2})(\bX - \bmustr_{\bX_2})\trans \omega(\bX)|T=2\}Var(\bX|T=2)^{-1}\bC_1 \\
&Cov( \varphi_i^{\mu_1}, \varphitld_i^{\balph_1}) = -P(T=2)^{-1}\bC_1\trans Var(\bX|T=2)^{-1} \\
&\quad E\{ (Y(1)-\mustr_1)(\bX-\bmustr_{\bX_2})\omega(\bX) | T=2\}, 
\end{eq*}
where $\bC_1 = Cov\{ Y(1),\bX|T=2\}$.  Moreover, the terms contributed for estimating $\bmu_{\bX_2}$ from \eqref{e:sig2} can be simplified as:
\begin{eq*}
\label{e:modvar}
&Var( \varphitld_i^{\bmu_{\bX_2}}) = P(T=2)^{-1}\bC_1\trans Var(\bX|T=2)^{-1}\bC_1 \\
& Cov(\varphi_i^{\mu_2}, \varphitld_i^{\bmu_{\bX_2}}) = -P(T=2)^{-1} \bC_1\trans Var(\bX|T=2)^{-1}\bC_2,
\end{eq*}
where $\bC_2 =Cov\{ Y(2),\bX|T=2\}$.
\end{lemma}
As long as there are no strong interactions between $\bX$ and the treatment, it can be expected that $\bC_1 \approx \bC_2$. In this case, if additionally the trial assignment model is at least approximately correctly specified, then $Var( \varphitld_i^{\bmu_{\bX_2}}) + 2Cov( \varphi_i^{\mu_2}, \varphitld_i^{\bmu_{\bX_2}}) \approx -\bC_1\trans Var(\bX|T=2)\bC_1 < 0$. Omitting the contributions from estimating $\bmu_{\bX_2}$ when estimating $\sigma^2$ thus tends to produce conservative standard errors. By bounding $\omega(\bX)$, a similar phenomenon occurs for terms contributed from estimating $\balph_1$ so that $Var(\varphitld_i^{\balph_1}) + 2Cov(\varphi_i^{\mu_1},\varphitld_i^{\balph_1}) < 0$ when the trial assignment model is correct and the lower and upper bounds for $\omega(\bX)$ are not too extreme.
These results suggest that omitting contributions for estimating both $\balph_1$ and $\bmu_{\bX_2}$ can yield conservative standard errors in scenarios where the trial assignment model is correctly specified.

A simple approach to estimating $\sigma^2$ is thus to estimate only $Var(\varphi_i^{\mu_1})$ and $Var(\varphi_i^{\mu_2})$, fully omitting contributions for $\balph_1$ and $\bmu_{\bX_2}$, as in:
\begin{eq}
\sigmahat^2_{fo} = \Varhat\left\{\varphi_i^{\mu_1}(\muhat_1,\balphhat_1)\right\} + \Varhat\left\{\varphi_i^{\mu_2}(\Delthat,\muhat_1)\right\},
\end{eq}
where $\muhat_1$ is the weighted average from the IPD as in the first term of \eqref{e:deltdef}, $\Varhat\left\{\varphi_i^{\mu_1}(\muhat_1,\balphhat_1)\right\}$ is the sample variance of $\varphi_i^{\mu_1}(\muhat_1,\balphhat_1)$, and $\Varhat\left\{\varphi_i^{\mutld_2}(\Delthat,\muhat_1)\right\} = S^2_{Y,22}/(N_{22}/N)$. Previous approaches to estimating standard errors for MAIC using robust sandwich estimators \citep{signorovitch2010comparative,phillippo20162016}, which are sometimes utilized in practice, are similar to this approach in that they ignore the variability from estimating $\balph_1$ and $\bmu_{\bX_2}$. 
Instead of fully ignoring the contributions for both $\balph_1$ and $\bmu_{\bX_2}$, another approach is to partially omit only the contribution for $\bmu_{\bX_2}$, as in:
\begin{eq}
&\sigmahat^2_{po} = \Varhat\left\{\varphi_i^{\mu_1}(\muhat_1,\balphhat_1)+ \varphitld_i^{\balph_1}(\balphhat_1,\muhat_1,\bmuhat_{\bX_2})\right\} + \Varhat\left\{\varphi_i^{\mu_2}(\Delthat,\muhat_1)\right\}, 
\end{eq}
where $\bmuhat_{\bX_2} = \bXbar_2$.  This is still feasible as $\varphitld_i^{\balph_1}$ does not involve IPD from the AGD trial. A final approach is to attempt to fully estimate $\sigma^2$.  Without any further assumptions, $Var(\varphitld_i^{\bmu_{\bX_2}})$ can be approximated by:
\begin{eq*}
&V^{\bmu_{\bX_2}} = -\frac{E\{ e^{\balphstrtrans_1(\bX_i-\bmustr_{\bX_2})}I(T_i=1)\}}{J^{\mu_1}(\balphstr_1)^2P(T_i=2)} \bCtld_{1}(\balphstr_1)\trans\bJ^{\balphstr_1}(\balphstr_1,\bmustr_{\bX_2})^{-1}\bCtld_{1}(\balphstr_1),
\end{eq*}
which partially simplifies $Var(\varphi_i^{\bmutld_{\bX_2}})$ under correct trial selection model to obviate the need for IPD from the AGD trial. The covariance term $Cov(\varphi_i^{\mu_2},\varphitld_i^{\bmu_{\bX_2}})$ can be bounded by the Cauchy-Schwartz inequality. This suggests estimating $\sigma^2$ by:
\begin{eq}
&\sigmahat^2_{cs} = \widehat{Var}\left\{\varphi_i^{\mu_1}(\muhat_1,\balphhat_1)+ \varphitld_i^{\balph_1}(\balphhat_1,\muhat_1,\bmuhat_{\bX_2})\right\} + \Varhat\left\{\varphi_i^{\mu_2}(\Delthat,\muhat_1)\right\} \\
&\qquad + \Vhat^{\bmu_{\bX_2}} + 2\left[\Varhat\left\{\varphi_i^{\mu_2}(\Delthat,\muhat_1)\right\} \Vhat^{\bmu_{\bX_2}}\right]^{1/2},
\end{eq}
where $\Vhat^{\bmu_{\bX_2}}$ is an empirical version of $V^{\bmu_{\bX_2}}$. Among these proposed approaches, we expect $\sigmahat_{fo}^2$ to be more conservative than $\sigmahat_{po}^2$ as it potentially omits negative contributions to the asymptotic variance when underlying assumptions are satisfied. We will see in the simulation results of Section \ref{s:sim} that this conservativeness of $\sigmahat_{fo}^2$ tends to improve its accuracy in approximating the true standard error in small samples, when both $\sigmahat^2_{fo}$ and $\sigmahat^2_{po}$ tend to underestimate, without paying a large price in terms of overestimation in large samples.  Such underestimation in small samples has also been observed for sandwich variance estimators in other settings\citep{kauermann2001note,fay2001small}. $\sigmahat_{cs}^2$ can be expected to be the most conservative, as it uses a very conservative bound for covariance. This estimator could be potentially useful in situations when conservative confidence intervals and hypothesis tests are prioritized over efficiency considerations.

\section{Simulations}
\label{s:sim}
We performed simulations to assess the finite sample bias of MAIC and alternative estimators.  In particular, we sought to identify scenarios where proposed approaches fail to provide reliable inferences. We also assessed the coverage and relative length of CIs based on the proposed variance estimators.  Besides the estimator $\Delthat$ (MAIC-NAB), which is based on \eqref{e:id}, we also consider an anchored MAIC approach (MAIC-ACB) based on \eqref{e:idalt}, $\Delthat^{ACB} = \Delthat - \{\sum_{i=1}^N I(Z_i=0,T_i=1)\omega(\bX_i;\balphhat_1)Y_i / \sum_{i=1}^N I(Z_i=0,T_i=1)\omega(\bX_i;\balphhat_1) - \Ybar_{20}\}$.
Additionally, we consider the method of \cite{bucher1997results} (BUC) and the basic formulation of simulated treatment comparisons of \cite{ishak2015simulation} (STC).  To be consistent in the scale of contrast with other methods, we implement BUC on the risk difference scale as $\Delthat^{BUC} = (\Ybar_{11} - \Ybar_{10}) - (\Ybar_{22} - \Ybar_{20})$.  As such, BUC would require that $E(Y|Z,\bX)$ be linear in $\bX$ to be unbiased, where $\bX$ are covariates that satisfy Assumption \ref{a:ignormod}.  We also considered simulations for contrasts on the log odds-ratio scale for all methods in Appendix \ref{apd:lgtsim}. For STCs, we assume a logistic regression model for $E(Y|Z=1,\bX)$:
\begin{eq*}
m_1(\bX;\bgamma) = g(\bgamma\trans\bXvec),
\end{eq*}
where $g(\cdot)$ denotes the inverse-logit link function for binary outcomes in the simulations.  This model is fit using data from active arm in the IPD trial, with the estimator denoted by $\bgammahat$.  It is then used to extrapolate the mean outcome had individuals in the AGD trial received treatment $Z=1$ by $\Delthat^{STC} = m_1(\bXbar_2;\bgammahat)-\Ybar_{22}$.  This approach would be unbiased if the model for $E(Y|Z=1,\bX)$ is correctly specified with $g(\cdot)$ being linear.  But even if $m_1(\bX;\bgamma)$ is correctly specified and $\bX$ satisfy the identification assumptions, still $E\{Y(1)|T=2\}=E\{ E(Y|Z=1,\bX)|T=2\} \neq E\{Y|Z=1,\bX=E(\bX|T=2)\} = m(\bmu_{\bX_2};\bgamma)$ when $g(\cdot)$ is non-linear, which results in bias.

We simulated data jointly for both the IPD and AGD trials in the case with binary $Y$ and continuous $\bX$.  In all scenarios, independent observations were simulated according to $\bX \sim N(\bzero, .8\bI_{p} +.2)$, $T|\bX \sim Ber\{P(T=2|\bX)\} + 1$, $Z \sim Ber(.5) \cdot T$, and $Y|\bX,Z,T \sim Ber\{ E(Y|\bX,Z,T)\}$, where:
\begin{eq}
\label{e:simmod}
logitP(T=2|\bX) &= \alpha_0 + \balph_1\trans\bX \\
logitE(Y|\bX,Z,T) &= \beta_0 +\{\bbeta_1\trans+\bbeta_3\trans I(Z>0)\}\bX + \beta_2 I(Z>0)+ I(Z=2)\beta_4,
\end{eq}
with $\alpha_0=0$ $\beta_0 = -1$, $\beta_2 = .1$, and $\beta_4=.5$.  For $P(T=2|\bX)$, $\balph_1 \neq \bzero$ induces imbalance in the distribution of $\bX$ in the IPD and AGD trial populations.  Any imbalance in a subvector of $\bX$ that also has a non-zero coefficient in $\bbeta_3$ subsequently induces confounding when comparing outcomes between trials.  $\bbeta_3 \neq \bzero$ results in treatment effect heterogeneity on the logit scale between active and placebo treatments.  We considered scenarios with no confounding, moderate confounding, and severe confounding, with the parameters:
\begin{eq*}
\emph{None: }&\balph_1 = (.25 \bone\trans_4,\bzero\trans_{p-4})\trans \quad \bbeta_1 = \bzero_{p} \quad \bbeta_3 = \bzero_{p}\\
\emph{Moderate: }&\balph_1 = (.25 \bone\trans_4,\bzero\trans_{p-4})\trans \quad \bbeta_1 = (.15 \bone\trans_4,\bzero\trans_{p-4})\trans \quad \bbeta_3 = (.1 \bone\trans_4,\bzero\trans_{p-4})\trans\\
\emph{Severe: }&\balph_1 = (.30 \bone\trans_4,\bzero\trans_{p-4})\trans \quad \bbeta_1 = (.25 \bone\trans_4,\bzero\trans_{p-4})\trans \quad \bbeta_3 = (.15 \bone\trans_4,\bzero\trans_{p-4})\trans.
\end{eq*}
This setup leads to a standardized mean difference \citep{cohen2013statistical} of approximately -.38 for the first 4 covariates 
and -.18 for the remaining covariates in the none and moderate settings and -.45 and -.22 in the severe setting.  To get a 
sense of the treatment effect heterogeneity, the standard deviation
of the true differences in the outcome probability for active versus placebo treatments in the IPD and AGD trials are 0 and 0 for 
the none setting, .05 and .07 for the moderate setting, and .07 and .09 for the severe setting.
We initially generated a large sample $\left\{ (Y_i,Z_i,T_i,\bX_i): i =1,\ldots,\Nstr\right\}$ and then randomly sub-sampled 
by arm in each trial, as described in 
Appendix \ref{apd:altsamp},
to include a fixed number of $n$ patients in all arms in the final sample.  We also provide arguments there to justify that
the proposed procedures for inference are still valid under this modified sampling scheme.

In each replicate of the data we calculated the four estimators for $\Delta$, as well as the four estimators for $\sigma^2$ discussed in Section \ref{ss:estvar}.  Besides $\sigmahat^2_{fo}$, $\sigmahat^2_{po}$, and $\sigmahat^2_{cs}$, we also implemented an estimator $\sigmahat^2_{sw}$ based on a sandwich estimator for regression coefficients when $\Delthat$ is implemented through a weighted linear regression, using the \texttt{sandwich} package in R with default options \citep{zeileis2004econometric}. The approach has been previously considered for $\Delthat^{ACB}$ \citep{phillippo20162016} and is essentially a direct calculation of the variance of $\Delthat$, treating the weights $\omega(\bX_i)$ and treatment assignment $T_i$ as fixed and plugging in estimators for the variance of $Y_i$ that allow for heteroskedasticity.
For benchmarking purposes, we also calculated an estimator of $\sigma^2$ using the full influence function, the estimator that would typically be computed had the IPD from all trials been available.  
The true $\Delta$ was calculated through simulating a large sample and calculating the mean difference of counterfactual outcomes had patients in the AGD population received treatment $Z=1$ and $Z=2$, according to $E(Y|\bX,Z,T)$ in \eqref{e:simmod}.  The percent bias for each estimator was then calculated as $R^{-1}\sum_{r=1}^R(\Delthat^{(r)} - \Delta)/\Delta$, where $\Delthat^{(r)}$ denotes an estimator calculated from data in the $r$-th replicate.  The bias simulations were repeated over the different confounding scenarios for sample sizes ranging $n=25$ to $n=500$ per arm and $p=5,10,15$.  
The CI coverage was calculated as $R^{-1}\sum_{r=1}^R I[\Delta \in \{\Delthat^{(r)} \pm z_{.975} N^{-1/2}\sigmahat^{(r)}\} ]$, where $z_{.975}$ denotes the $.975$ quantile of a standard normal and $\sigmahat^{(r)}$ denotes an estimator of the asymptotic variance estimated from data observed in the $r$-th replicate.  Relative CI length was calculated as $R^{-1}\sum_{r=1}^R \sigmahat^{(r)}/\sigmahat_{emp}$, where $\sigmahat_{emp}^2=(R-1)^{-1}\sum_{r=1}^R \{ \Delthat^{(r)} -R^{-1}\sum_{r=1}^R \Delthat^{(r)}\}^2$ is the empirical variance of $\Delthat^{(r)}$ over all repetitions.
The coverage simulations were conducted under the moderate confounding scenario for $n=25$ to $n=500$ per arm with $p=5,15$.  For each set of simulations we ran $R=5,000$ replicates to well approximate the tail probabilities when evaluating CI coverage.

\subsection{Simulation Results}
\begin{figure}[h!]
  \caption{Percent bias of estimators by degree of confounding, sample size per arm ($n$), and number of covariates ($p$).}
  \label{f:bias}
\begin{center}
  \includegraphics[scale=0.0705]{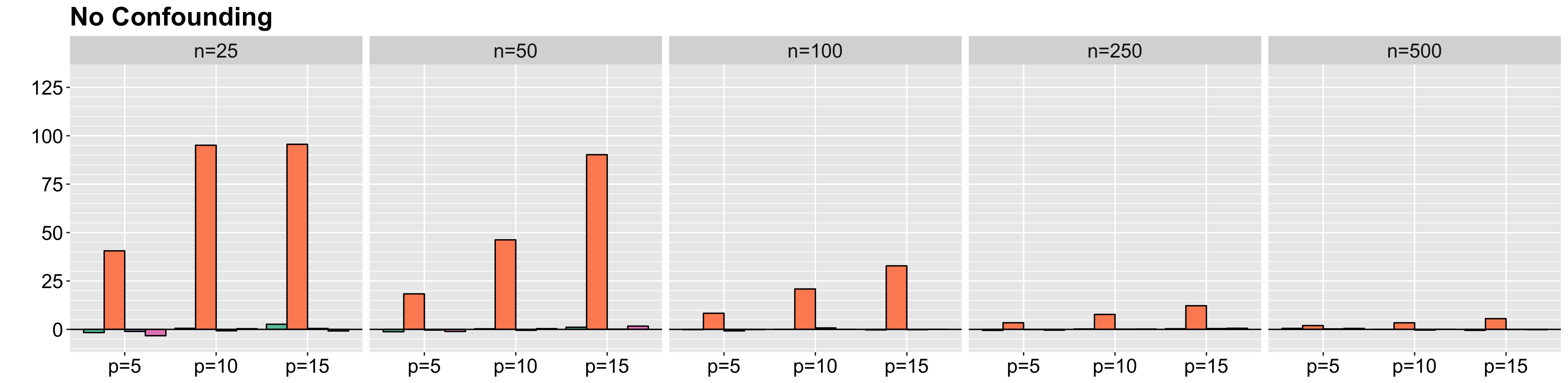} 
  \includegraphics[scale=0.0705]{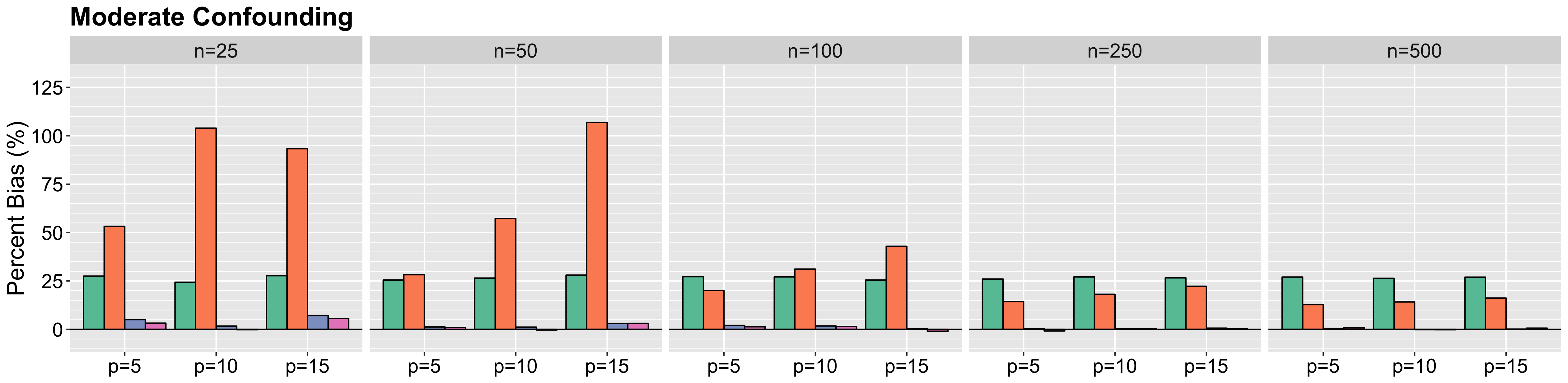}
  \includegraphics[scale=0.0705]{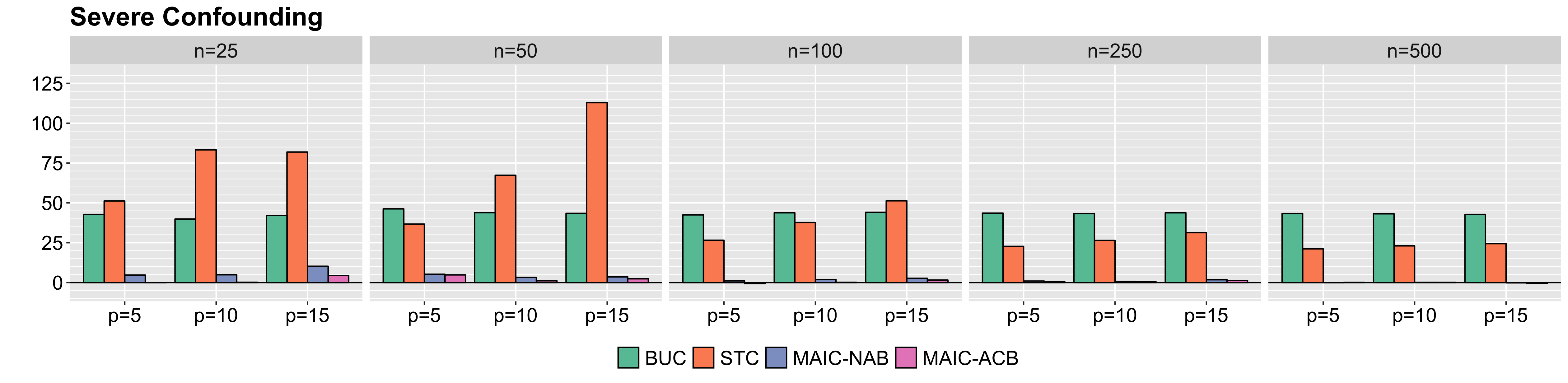}
\end{center}
BUC: Method of Bucher et al, STC: Simulated treatment comparison, MAIC-NAB: Non-anchor-based MAIC, MAIC-ACB: Anchor-based MAIC.
\end{figure}

The results for the bias simulations are presented in 
Figure \ref{f:bias}.
Both MAIC estimators generally exhibited negligible bias across the scenarios considered.  The bias of BUC increases with the degree of confounding and results from the non-linear link function for $E(Y|\bX,Z,T)$ and interaction between $\bX$ and $Z$ in \eqref{e:simmod}.  Similar results hold even when considering contrasts on the log odds-ratio scale in Appendix \ref{apd:lgtsim}. STC also incurred bias that increases with the degree of confounding due to the non-linear link.  STC, however, generally had lower bias than BUC for larger $n$.  Extrapolation based on fitting $m_1(\bX;\bgamma)$ appears to outperform placebo adjustment in BUC in large samples.  The bias of STC in small samples also appears to be more sensitive to increasing $p$ than MAIC and BUC.

\begin{figure}[h!]
\caption{Coverage and relative length of 95\% CI's for $\Delthat$ by size per arm ($n$) and number of covariates ($p$) in the moderate confounding scenario. Some points and bars for $n=25$ are omitted if they take values beyond the axis limits. $\sigmahat_{fo}^2$ and $\sigmahat_{sw}^2$ achieve close to nominal coverage for $n \geq 50$ and only slightly conservative in length.}
\label{f:cov}
\begin{center}
\includegraphics[scale=.31]{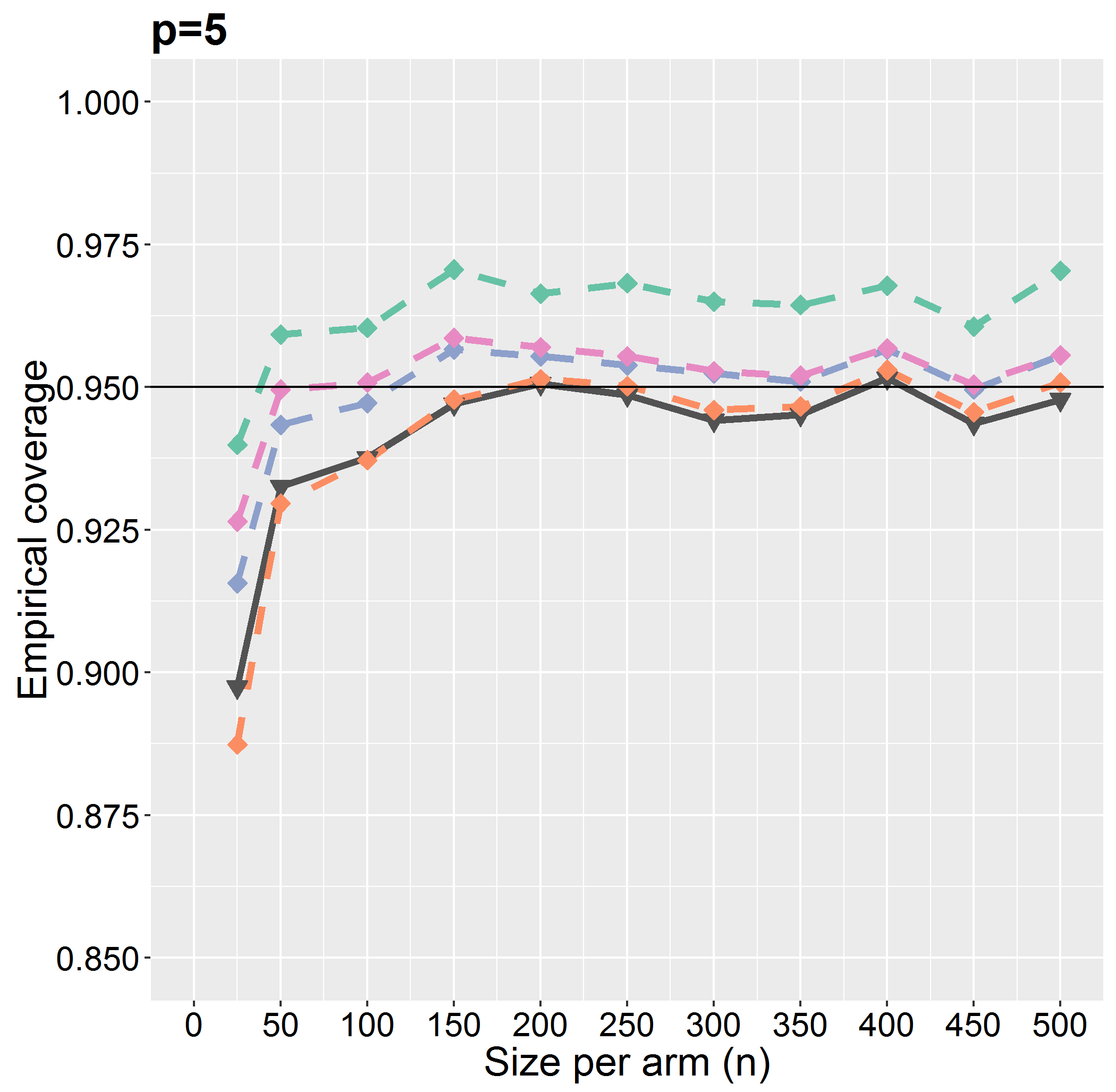} 
\includegraphics[scale=.31]{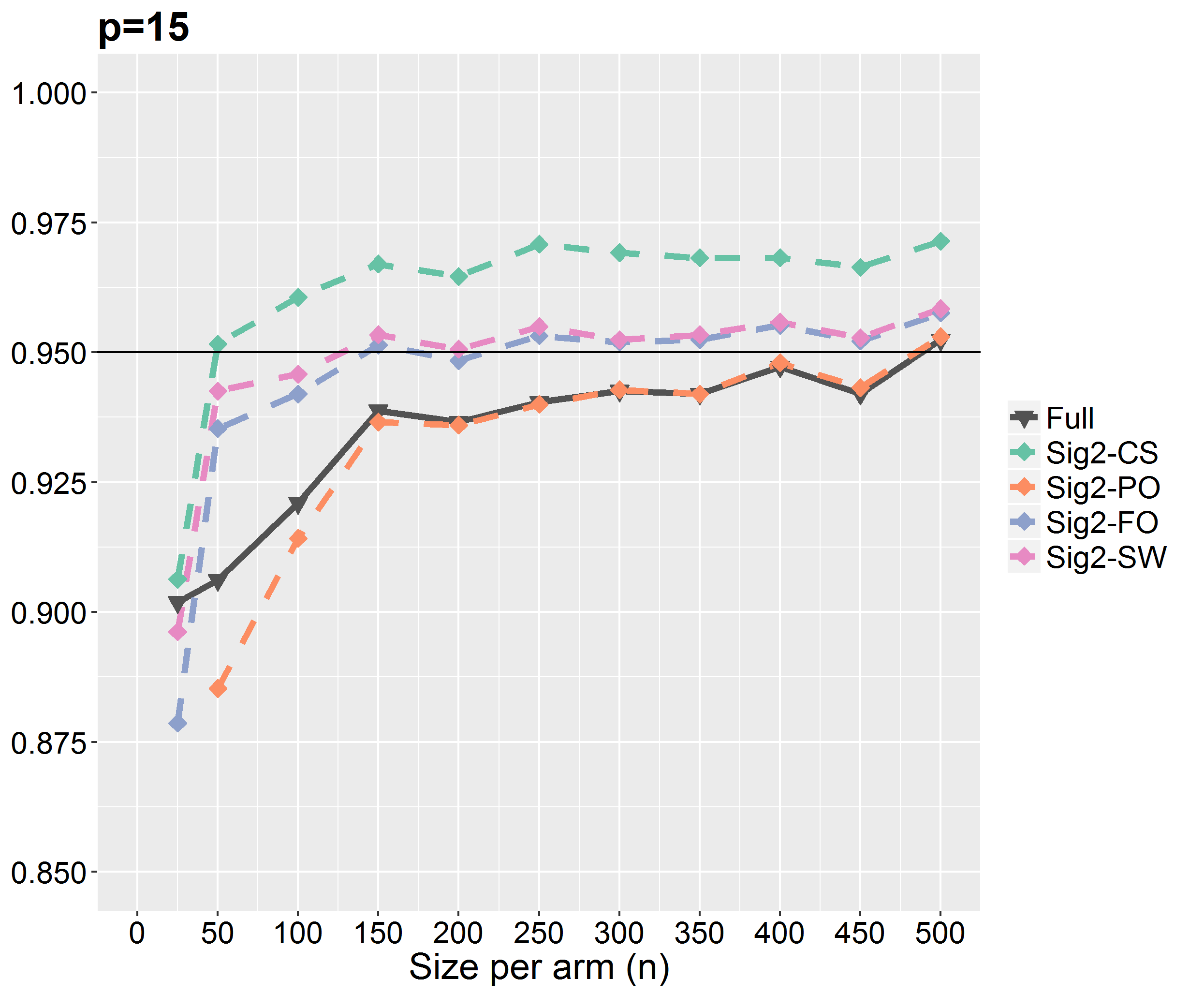}
\includegraphics[scale=.29]{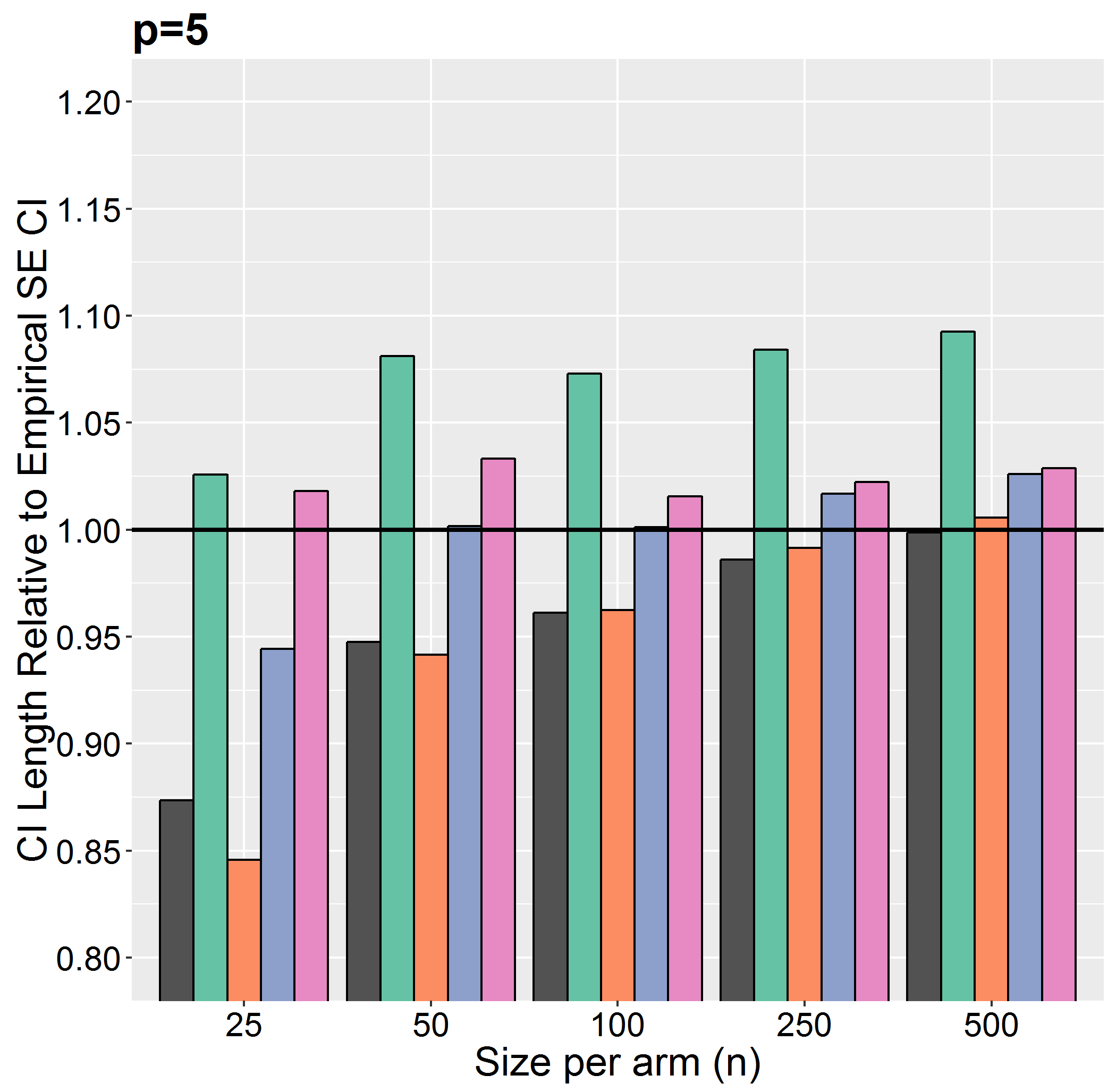} 
\includegraphics[scale=.29]{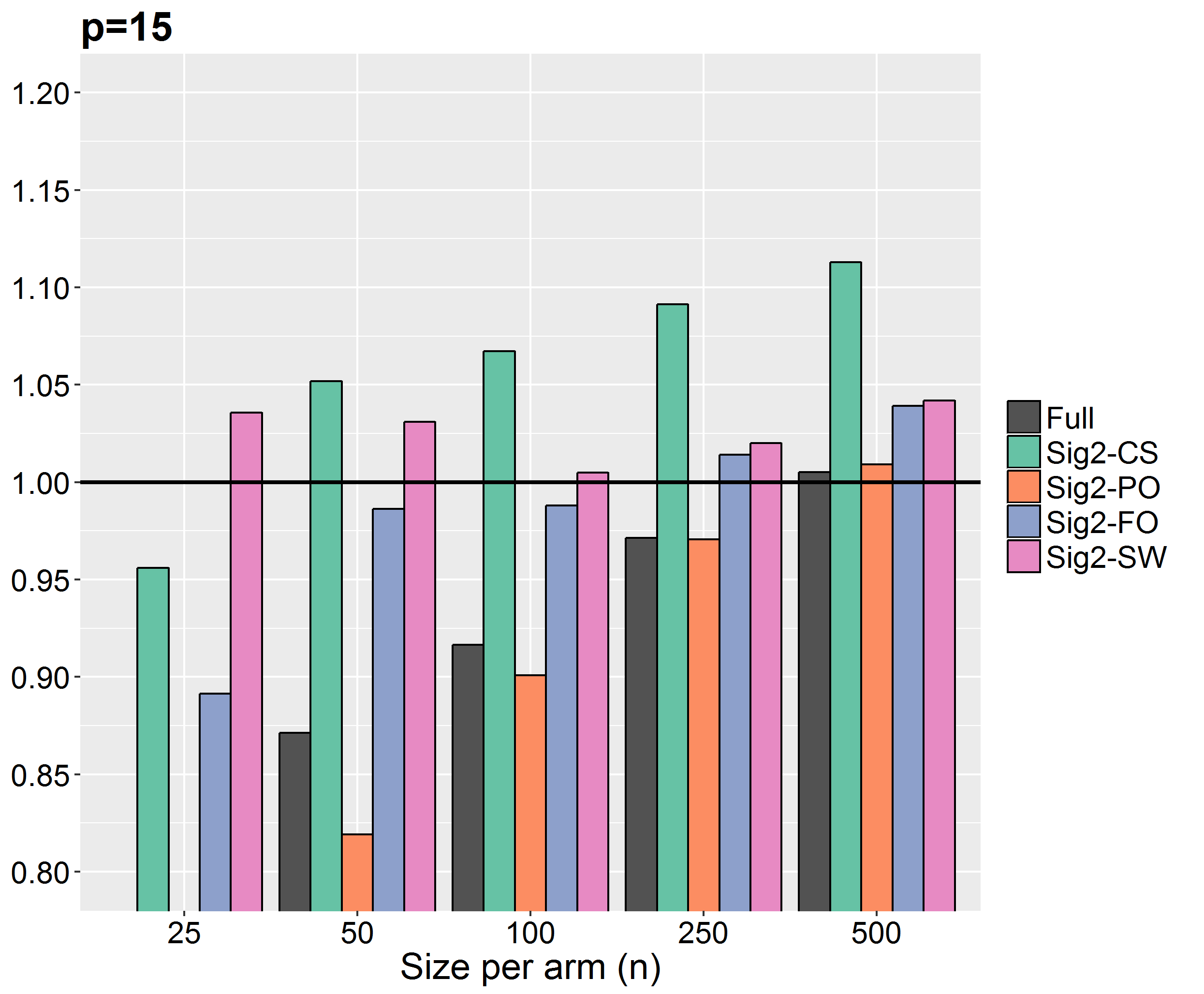}

Full: Based on full influence function, Sig2-CS: Based on $\sigmahat^2_{cs}$, Sig2-PO: Based on $\sigmahat^2_{po}$, Sig2-FO: Based on $\sigmahat^2_{fo}$, Sig2-SW: Based on $\sigmahat^2_{sw}$. 
\end{center}
\end{figure}

Results from the coverage simulations are presented in Figure \ref{f:cov}.  CIs based on $\sigmahat^2_{fo}$ and $\sigmahat^2_{sw}$ generally achieve close to nominal coverage for $n\geq 50$ per arm. They are only slightly conservative in terms of both coverage and length, relative to empirical estimates, in large samples. Approximating the behavior of CIs based on the full influence function, CIs based on $\sigmahat^2_{po}$ also achieve close to nominal coverage for $n\geq 150$ per arm when $p=5$ but exhibits slight undercoverage for smaller $n$ and when $p=15$. This could be related to underestimation observed for sandwich estimators in small samples as discussed in Section \ref{ss:estvar} and can be expected with larger number of parameters. Still, in large samples both still show excellent performance. For all the approaches considered, there appears to be a drop in coverage for $n=25$. CIs for studies with such small samples should be interpreted with caution. The CI based on $\sigmahat^2_{cs}$ is consistently conservative, achieving coverage of around 97\% in most scenarios and exhibiting a length around 5-10\% longer than CIs based on empirical estimates of the standard error.

In the Supplementary Materials, we also provide an extended set of simulation results on both bias and coverage for a range 
of data settings, including for both binary and continuous $Y$, to further elucidate when MAIC and its standard 
error estimators are or are not reliable. 

\section{Data Application}
\label{s:application}
Duchenne muscular dystrophy (DMD) is a rare neuromuscular disorder in which patients experience progressive
muscle degeneration that often leads to loss of ambulation during adolescence, respiratory and cardiac dysfunction in
early adulthood, and eventually premature mortality \citep{emery2015duchenne}.
Due to the rarity of the disease, recruitment of DMD patients into clinical trials can be challenging.  There is 
strong interest in adopting single-arm designs for future DMD trials and using natural history (NH) data for external controls. 
Such external control groups have previously informed drug approval in other rare diseases \citep{guideline2000choice,us2006fda,us2017fda}.  
However, indirect comparisons to external controls is subject to the risk of bias due to potential differences in patient 
baseline characteristics, along with other factors such as those in Table \ref{t:desgnbias}.

To illustrate MAIC, we considered a ``negative control’’ study to assess whether a NH cohort from a prospective non-interventional 
study (DMD-PRO-01), provided by CureDuchenne, a 501(3)c patient foundation, are sufficiently comparable to the placebo arm
of a phase III DMD trial for tadalafil (PBO), provided by Eli Lilly, using IPD from DMD-PRO-01 and AGD from the PBO trial. If the NH setting is sufficiently comparable to the trial, 
mean outcomes between the studies should be similar after adjustment for differences in observed characteristics. The data were accessed through 
the Collaborative Trajectory Analysis Project (cTAP), a collaboration aiming to improve clinical trial design and interpretation in DMD.

We focused on a binary outcome for clinically significant worsening of the North Star Ambulatory 
Assessment (NSAA), defined as decrease of $\geq 3$ units from baseline to 
week 48 after study initiation \citep{ricotti2016northstar}.  
We implemented MAIC-NAB, in addition to STC and a naïve comparison that directly contrasted mean outcomes between the two studies 
without adjustment for baseline characteristics.  Comparisons 
were conducted on the log odds-ratio scale.  For instance, MAIC-NAB estimates are obtained as $\Delthat_g = g(\muhat_1)-g(\Ybar_{22})$ 
with $g(u)=log\{ u/(1-u)\}$.  For MAIC and STC, we include baseline characteristics from Table \ref{t:blchar} in $\bX$.  
These characteristics were selected based on studies evaluating prognostic factors for ambulatory outcomes in DMD and 
were thought to be important confounding factors \citep{mazzone2016timed,goemans2016individualized}.  Data on other potential 
confounding factors such as genetic markers were not available from both studies, and interpretations of the results should 
bear in mind that other unknown or unobserved factors could still bias the results after adjustment.
Estimators of the asymptotic variance on the log-odds ratio scale based on strategies discussed in Section \ref{ss:estvar} were obtained 
(detailed in Appendix \ref{apd:nonlin}). We report SEs based on $\sigmahat_{fo}^2$, which we chose since the sample size is not small. 
For the naïve comparison, we use the usual SE estimator for log odds-ratios based on the delta method \citep{bland2000odds}. 
Only the point estimate is available for STC.
Two-sided p-values were also reported 
from Wald tests for the null hypothesis that there is no difference in the proportion of patients with NSAA worsening between studies.  Comparisons 
were repeated with and without application of the inclusion/exclusion criteria from the tadalafil trial to DMD-PRO-01 
(age 7-14 years, steroid duration $\geq$ 6 months, baseline 6 min walk distance 200-400 meters).

Table \ref{t:blchar} reports the key characteristics available in both cohorts.  Patients in the NH cohort, on average, were 
younger, had better ambulatory function and rise time, and shorter duration of steroid treatment.  The indirect comparison 
results are reported in Table \ref{t:icres}.  A naïve comparison of outcomes suggests 
that there were significant differences in NSAA worsening, with odds of worsening about 40\% 
lower in NH vs. PBO (OR=0.58; 95\% CI: 0.34 to 0.99, $p=0.04$).  After MAIC adjustment, the magnitude of differences attenuated 
and was no longer statistically significant (OR = 1.14, 95\% CI: 0.64 to 2.04, $p=0.66$).  The standard error 
increased slightly on the log odds scale from 0.27 to 0.30 after adjustment with MAIC.  Results were similar for STC (OR=1.20).  
Findings were also similar after applying the trial inclusion/exclusion criteria.  These results suggest that NH data is comparable
to trial data after appropriate adjustment for baseline characteristics.

\begin{table}[htbp]
  \centering
  \caption{Baseline characteristics of patients from DMD-PRO-01 (NH) and tadalafil trial (PBO).}
    \begin{tabular}{lcc}
    \toprule
    \textbf{Baseline Characteristics} & \textbf{NH (n=152)} & \textbf{PBO (n=90)} \\
    \midrule
    Age (years) & 8.8   & 9.3 \\
    NSAA score & 24.4  & 22.6 \\
    Six min walk distance (meters) & 374.3 & 348.5 \\
    Rise time $\geq$ 5 seconds & 50.0\% & 74.4\% \\
    Steroid duration $\geq$ 12 months & 70\%   & 90\% \\
    Height (cm) & 122.1 & 125.5 \\
    Weight (kg) & 28.2  & 30.6 \\
    \bottomrule
    \end{tabular}%
  \label{t:blchar}%

  NSAA: North Star Ambulatory Assessment
\end{table}%

\begin{table}[htbp]
  \centering
  \caption{Indirect comparison of NSAA worsening between NH vs. PBO cohorts.}
    \begin{tabular}{ccccccc}
    \toprule
          & \multicolumn{1}{c}{\textbf{Method}} & \multicolumn{1}{c}{\textbf{OR}} & \multicolumn{1}{c}{\textbf{95\% CI}} & \multicolumn{1}{c}{\textbf{Log OR}} & \textbf{SE} & \textbf{p-value} \\
    \midrule
    \multirow{3}[1]{*}{\pbox{5cm}{\vspace{-.2cm}No inclusion/exclusion \\ criteria applied}} & Naive & 0.58  & (0.34, 0.99) & -0.54 & 0.27  & 0.04 \\
          & STC   & 1.20   & -     & 0.19  & -     & - \\
          & MAIC-NAB  & 1.14  &  (0.64, 2.04) & 0.13  & 0.3   & 0.66 \\ \hline
    \multirow{3}[1]{*}{\hspace{.3cm}\pbox{5cm}{\vspace{-.15cm}With inclusion/exclusion \\ criteria applied}} & Naive & 1.78  & (0.95, 3.34) & 0.58  & 0.32  & 0.07 \\
          & STC   & 1.75  & -     & 0.56     & -     & - \\
          & MAIC-NAB  & 1.42  & (0.69, 2.92) & 0.35  & 0.37  & 0.34 \\
    \bottomrule
    \end{tabular}%
    \label{t:icres}%

    Naive: Unadjusted comparison, STC: Simulated treatment comparison, MAIC-NAB: Non-anchor-based MAIC.
\end{table}%

\section{Discussion}
\label{s:discuss}

As with any nonrandomized treatment comparison, indirect comparisons through MAIC can be biased by differences in unobserved confounders.  Researchers ought to include covariates $\bX$ such that Assumptions \ref{a:ignor} or \ref{a:ignormod} are satisfied as much as possible and recognize the limitations of indirect comparison when confounders are unobserved.  In the literature on estimating average treatment effects, it is well-known that including covariates associated with only the outcome increases efficiency, while covariates associated with only the exposure decreases efficiency \citep{lunceford2004stratification,brookhart2006variable, rotnitzky2010note,de2011covariate}. The same results are expected to hold when estimating the weights for MAIC, viewing trial selection as the ``exposure.''  It is thus generally advisable to include in $\bX$ observed covariates that are known or suspected to be associated with outcomes, even if their imbalance between trials is minor \citep{rubin1996matching,lunceford2004stratification,stuart2010matching}.   As discussed in Section \ref{ss:ident}, with anchor-based MAIC, there are cases where it is not necessary to adjust for covariates in $\bX$ that are not effect modifiers. However, it is generally difficult to be certain that any covariate is not an effect modifier, especially for novel therapies. A simple objective approach to covariate selection is to include all covariates available for both the IPD and AGD that cannot be ruled out as having no associations with outcomes. Known prognostic covariates that are unobserved should be noted among the limitations. Sensitivity analyses on the impact of adding or removing covariates from $\bX$ can also be informative.

Even when all confounders are observed, MAIC still relies on the trial assignment model from \eqref{e:trlmod} being at least approximately correct for a given $\bX$ to make appropriate adjustments.  It is important for researchers to consider its specification, such as whether higher-order polynomial or interaction terms are plausible and the corresponding requisite AGD, such as standard errors or correlations, are available.  Other parametric models besides logistic regression could also be used, but the estimating equation that balances trial covariates, as in \eqref{e:momcnd},  must admit a unique solution for the model parameters.  

Beyond unobserved confounding and model specification, Assumption \ref{a:posit} is also an important assumption.  When $P(T=1|\bX=\bx)$ is close to $0$ for some $\bx \in \Xscr$, there exist some subpopulation in the AGD population that does not have strong overlap with the IPD population.  This leads some observations in the IPD to receive extreme weights and dominate the re-weighted sample, resulting in poor performance in point and standard error estimation.  A useful diagnostic tool for lack of overlap is the effective sample size, defined by $\Ntld_{1z}=\{ \sum_{T_i=1,Z_i=z}\omega(\bX_i;\balphhat_1)\}^2/\sum_{T_i=1,Z_i=z}\omega(\bX_i;\balphhat_1)^2$ for weighted samples from arm $z=0,1$ of the IPD.  $\Ntld_{1z}$ is an approximation from importance sampling that down-weights the sample size by the approximate relative efficiency between a sample average using data from a target distribution and a weighted average from a proposal distribution \citep{kong1992note}.  Violations or near-violations of Assumption \ref{a:posit} would tend to inflate some $\omega(\bX_i;\balphhat_1)$ and deflate $\Ntld_{1z}$, signaling poor overlap. { In simulations, coverage in cases when $\Ntld_{11}\leq p$ ranged 68-88\% for CIs based on $\sigmahat^2_{fo}$, in the small sample scenario with $n=25$, $p=5,15$, and moderate confounding. This suggests that inferences based on $\Delthat$ are suspect in real datasets when $\Ntld_{11}$ is less than or close to $p$.}  While $\Ntld_{1z}$ is termed the ``effective sample size'' for arm $z=0,1$, it has no direct bearing on statistical inferences, including standard error and CI estimation, and should be viewed as a rough indicator only. Poor performance in estimation can also result when $p$ is large or covariates in $\bX$ are heavily correlated, which yields large standard errors for $\balphhat_1$.  One possible remedy would be to add a regularization term to \eqref{e:alphee}.  

As illustrated in Section \ref{s:application}, it is possible to consider a non-definitive test for the adequacy of Assumptions \ref{a:randm}-\ref{a:ignor} and specification of trial selection model using data from a common comparator arm, if available, provided the studies are deemed to be sufficiently similar in other aspects.  Let 
\begin{eq*}
\Delthat_0 &= \sum_{i=1}^N I(Z_i=0,T_i=1)\omega(\bX_i;\balphhat_1)Y_i/\sum_{i=1}^N I(Z_i=0,T_i=1)\omega(\bX_i;\balphhat_1) \\
&\qquad- \Ybar_{20}
\end{eq*}
denote the difference in outcomes under the common comparator after weighting.  Under the null that these assumptions hold, $N^{1/2}\Delthat_0 \overset{d}{\to}N(0,\sigma^2_0),$ where $\sigma^2_0$ is analogous to $\sigma^2$ for outcomes under the common comparator.  Consequently a test that rejects when $N^{1/2}|\Delthat_0|/\sigmahat_0 > z_{1-{\alpha/2}}$  is a level $\alpha$ test, where $\sigmahat^2_0$ is an estimator of $\sigma_0^2$ that can be constructed using similar strategies as in Section \ref{ss:estvar}.  The type I error may be conservative if a conservative strategy for estimating $\sigma^2_0$ is used.  Rejection in such a test suggests violation of some assumption, but failure to reject does not verify the assumptions.
For instance, there may be unobserved confounders that impact outcomes in the active treatment arms but not for the placebo arm.

MAIC enables estimation of causal contrasts between studies when IPD is available for one study and only AGD is available
for other studies.  This is a common occurrence for researchers who have access to data from their own study but only 
published AGD from other studies.  However, we are keen to emphasize that appropriate sharing of IPD is the preferred approach
when possible as it offers important advantages over settings where IPD is available only from some studies. 
  With the full IPD, the pooled data can 
be regarded as observational data for which the wide array of methods developed for causal inference can be applied.
In particular, while MAIC only enables estimation of causal contrasts in the AGD population, having the full IPD offers
additional flexibility in allowing for estimation of treatment contrasts in other target populations.  This not only
provide insight into treatment effect heterogeneity between populations but can also help circumvent issues with violations or
near-violations of study assumptions discussed above.  For example, if one study enrolled a more inclusive patient population
than the other, that study can be considered as the $T=1$ study so that it would be more plausible for Assumption \ref{a:posit} to
be satisfied.  Similarly, if one study collected a richer set of baseline prognostic covariates in $\bX$, that study can
be set as the $T=1$ study to more convincingly satisfy Assumption \ref{a:ignor}.
Access to full IPD also enables diagnostic assessments of the goodness of fit and calibration of the propensity score model,
which is a standard for propensity score analyses.

Another direction of future research will be to consider efficient estimation of $\Delta$.  { In particular, it would be of interest to consider whether any estimator with only AGD available in one trial can still achieve the known semiparametric efficiency bound for $\Delta$ \citep{hahn1998role}. Recently, entropy balancing, a method that estimates causal effects via weights that minimize the relative entropy with the distribution of covariates in the control population, has been shown to be locally semiparametric efficient if the logit of the propensity scores and mean outcomes in the treated population are both linear in the covariates $\bX$ \citep{zhao2017entropy}. 
Moreover, it was be also shown to be doubly robust in that it is consistent if either the logit propensity score or outcome model is linear in $\bX$.  
Since entropy-balancing coincides with $\Delthat$ when a logistic regression model is used for trial assignment, this immediately indicates that MAIC is also locally semiparametric efficient and doubly robust in the same sense. It would be of interest to consider whether doubly robust and efficient estimators are available for more general models when full IPD has been withheld for one treatment group. Developments of extensions for MAIC along these lines are underway.}

\appendix

\section{Inferences on Non-Linear Scale}
\label{apd:nonlin}
When the treatment effect on a different scale of contrast is of interest, $\Delta_g$ can be estimated by first estimating $E\{ Y(1)|T=2\}$ and $E\{ Y(2)|T=2\}$ and then applying a specified $g(\cdot)$ transformation.  The same considerations for identification and estimation of $E\{ Y(1)|T=2\}$ and $E\{ Y(2)|T=2\}$ applies as in Sections \ref{ss:ident} and \ref{ss:estimation}.  The estimator in this case would then be:
\begin{eq}
\Delthat_g = g(\muhat_1) - g(\Ybar_{22}),
\end{eq}
where $\muhat_1$ is the weighted average for treatment 1 using the IPD.  

Applying the delta method, when $g(u)$ is differentiable and non-zero valued at $u = \mu_1$ and $u  = \mu_2$, the influence function for $\Delthat_g$is given by:
\begin{eq*}
N^{1/2}(\Delthat_g - \Delta_g) &= N^{-1/2}\sum_{i=1}^N \psi_i^{\mu_2}(\Deltstr,\mustr_1) + \psi_i^{\mu_1}(\mustr_1,\balphstr_1) \\
&\qquad + \psitld_i^{\balph_1}(\balphstr_1,\mustr_1,\bmustr_{\bX_2}) + \psitld_i^{\bmu_{\bX_2}}(\balphstr_1,\mustr_1,\bmustr_{\bX_2}) + o_p(1),
\end{eq*}
where:
\begin{eq*}
&\psi_i^{\mu_2}(\Delta,\mu_1) = \varphi_i^{\mu_2}(\Delta,\mu_1)g'(\mu_1-\Delta)\\
&\psi_i^{\mu_1}(\mu_1,\balph_1)  = \varphi_i^{\mu_1}(\mu_1,\balph_1)g'(\mu_1)\\
&\psitld_i^{\balph_1}(\balph_1,\mu_1,\bmu_{\bX_2}) \\
&\qquad = J^{\mu_1}(\balph_1)^{-1}\bCtld_1(\mu_1,\balph_1)\trans\bJ^{\balph_1}(\balph_1,\bmu_{\bX_2})^{-1}\bU_i^{\balph_1}(\balph_1,\bmu_{\bX_2})g'(\mu_1)\\
&\psitld_i^{\bmu_{\bX_2}}(\balph_1,\mu_1,\bmu_{\bX_2}) \\
&\qquad = J^{\mu_1}(\balph_1)^{-1}\bCtld_1(\mu_1,\balph_1)\trans\bJ^{\balph_1}(\balph_1,\bmu_{\bX_2})^{-1}\bU_i^{\bmu_{\bX_2}}(\balph_1,\bmu_{\bX_2})g'(\mu_1)
\end{eq*}
are modifications of the original influence function with $g'(u) = \ddu g(u)$.  In parallel to \eqref{e:sig2}, the asymptotic variance of $\Delthat_g$ can be expressed as:
\begin{eq}
\label{e:sig2g}
\sigma_g^2 &= \{Var( \psi_i^{\mu_1})  + Var(\psi_i^{\mu_2})\} + \{Var(\psitld_i^{\balph_1}) +2Cov(\psi_i^{\mu_1},\psitld_i^{\balph_1}) \} \\
&\qquad + \{ Var(\psitld_i^{\bmu_{\bX_2}}) + 2Cov(\psi_i^{\mu_2},\psitld_i^{\bmu_{\bX_2}})\},
\end{eq}
where the arguments of the components of the influence function are suppressed but implicitly evaluated at their respective truth.
Based on similar considerations, the three proposed estimators for $\sigma^2_g$ are:
\begin{eq}
&\sigmahat^2_{g,fo} = \Varhat\{\psi_i^{\mu_1}(\muhat_1,\balphhat_1)\} + \Varhat\{\psi_i^{\mu_2}(\Delthat,\muhat_1)\} \\
&\sigmahat^2_{g,po} = \Varhat\{\psi_i^{\mu_1}(\muhat_1,\balphhat_1)\} + \Varhat\{\psitld_i^{\balph_1}(\balphhat_1,\muhat_1,\bmuhat_{\bX_2})\} + \Varhat\{\psi_i^{\mu_2}(\Delthat,\muhat_1)\} \\
&\sigmahat^2_{g,cs} = \Varhat\left\{\psi_i^{\mu_1}(\muhat_1,\balphhat_1)  + \psitld_i^{\balph_1}(\balphhat_1,\muhat_1,\bmuhat_{\bX_2})\right\} + \Varhat\{\psi_i^{\mu_2}(\Delthat,\muhat_1)\}\\
&\qquad\qquad +\Vhat^{\bmu_{\bX_2}}g'(\muhat_1)^2 + 2 [ \Varhat\{\psi_i^{\mu_2}(\Delthat,\muhat_1)\}\Vhat^{\bmu_{\bX_2}}g'(\muhat_1)^2]^{1/2}.
\end{eq}
As in Section \ref{ss:estvar}, $\sigmahat^2_{g,fo}$ tends to outperform $\sigmahat^2_{g,po}$ in small samples, whereas $\sigmahat^2_{g,cs}$ provides a conservative estimate.

\section{Simulation Results on Logit Scale}
\label{apd:lgtsim}
We repeated the simulations to assess the bias of the corresponding estimators modified to estimate the treatment effect on the logit scale, i.e. $\Delta_g$ with $g(u) = log\{u/(1-u)\}$.  This has previously been the recommended scale for binary outcomes HTA applications \citep{phillippo20162016}.  The performance of CIs for MAIC-NAB based on the modified estimators of the asymptotic variance were also assessed.  The same simulation settings as those in Section \ref{s:sim} were used throughout.  MAIC-ACB was calculated based on $\Delthat_g^{ACB} = \Delthat_g - g\{ \sum_{i=1}^N I(Z_i = 0,T_i=1)\omega(\bX_i;\balphhat_1)Y_i/\sum_{i=1}^N I(Z_i=0,T_i=1)\omega(\bX_i;\balphhat_1)\} + g(\Ybar_{20})$.

\begin{figure}[h!]
\caption{Coverage and relative length of 95\% CI's for $\Delthat$ by size per arm ($n$) and number of covariates ($p$) in the moderate confounding scenario. Some points and bars for $n=25$ are omitted if they take values beyond the axis limits.}
\label{f:covlogit}
\begin{center}
\includegraphics[scale=.31]{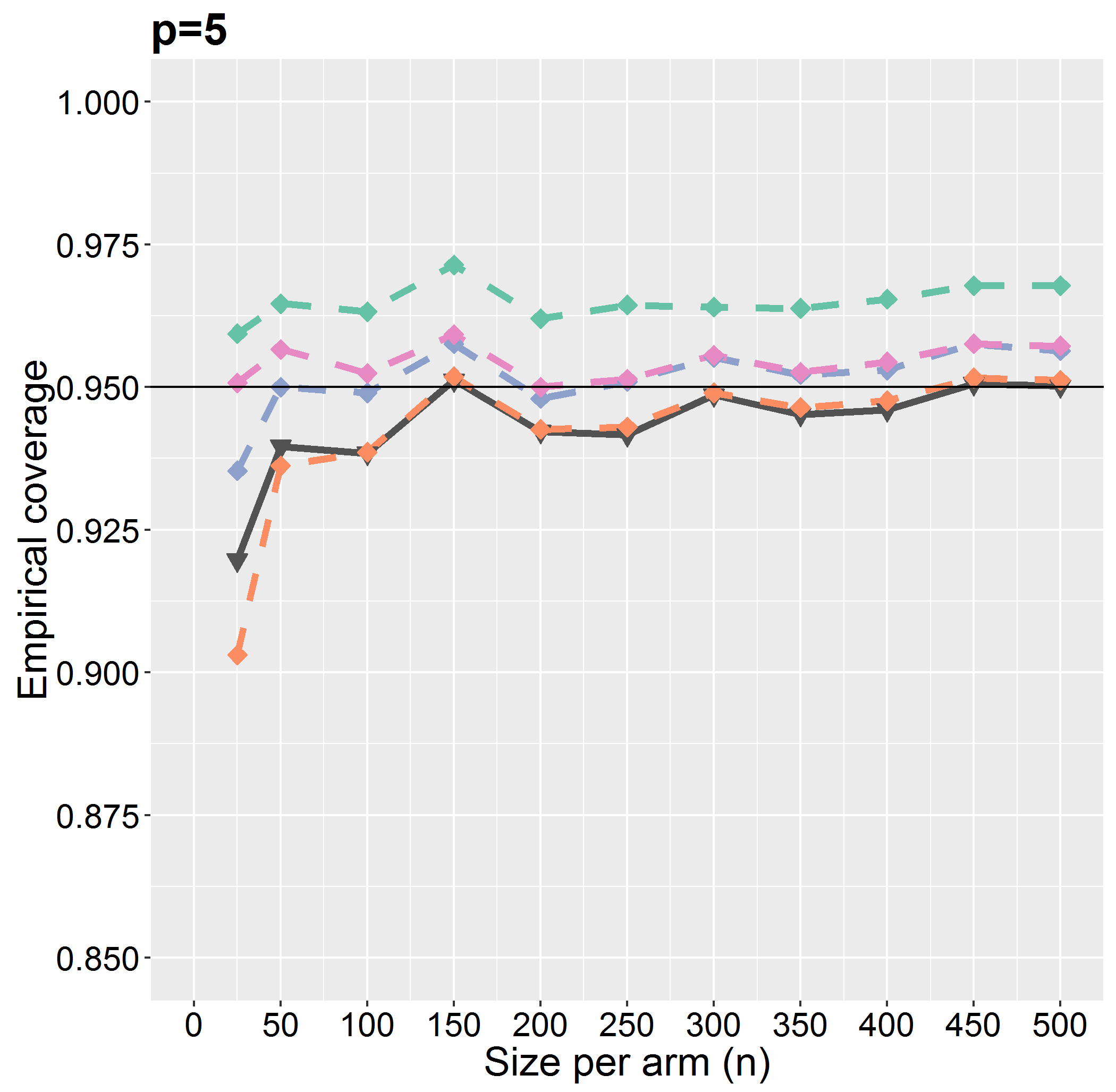} 
\includegraphics[scale=.31]{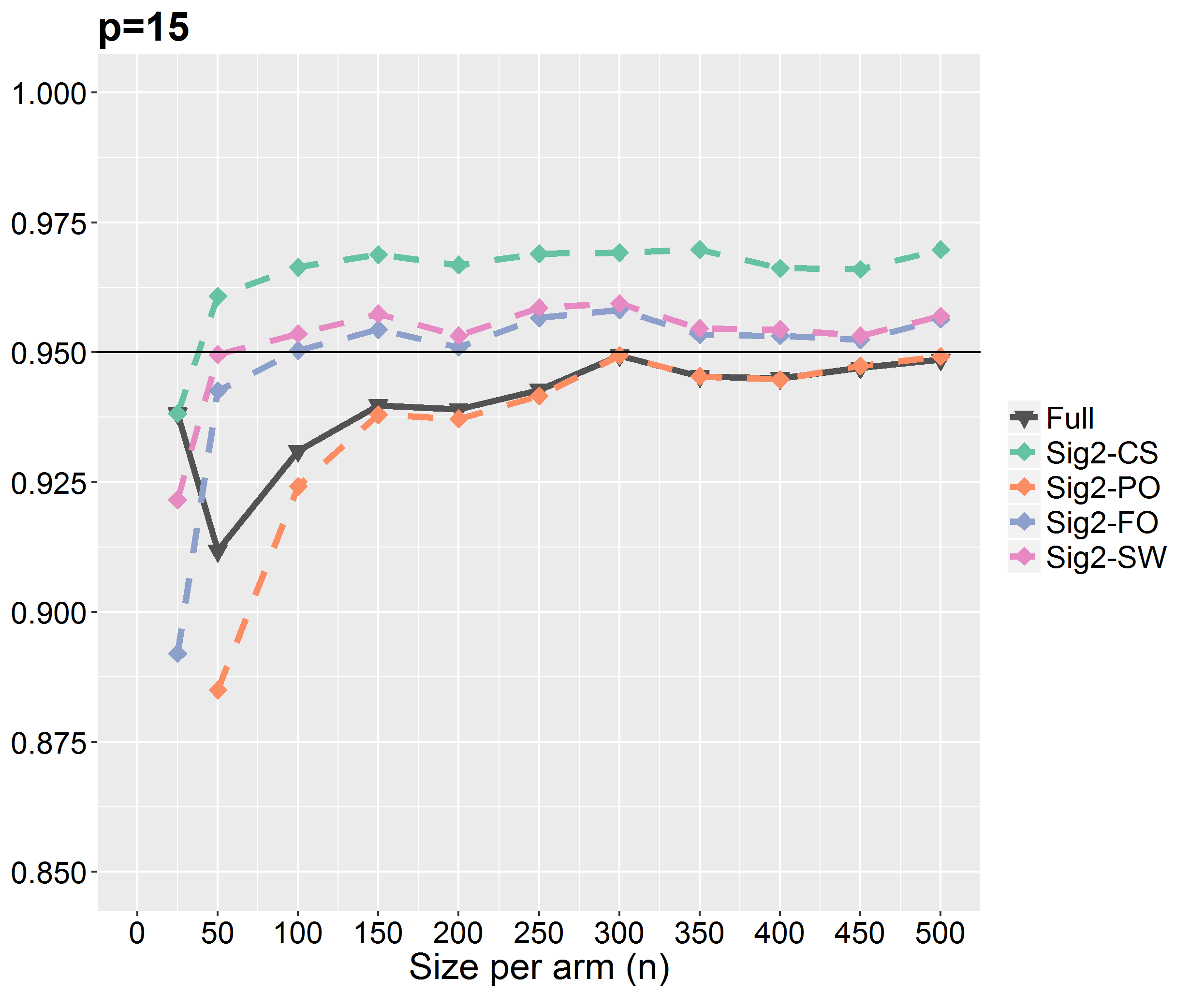}
\includegraphics[scale=.29]{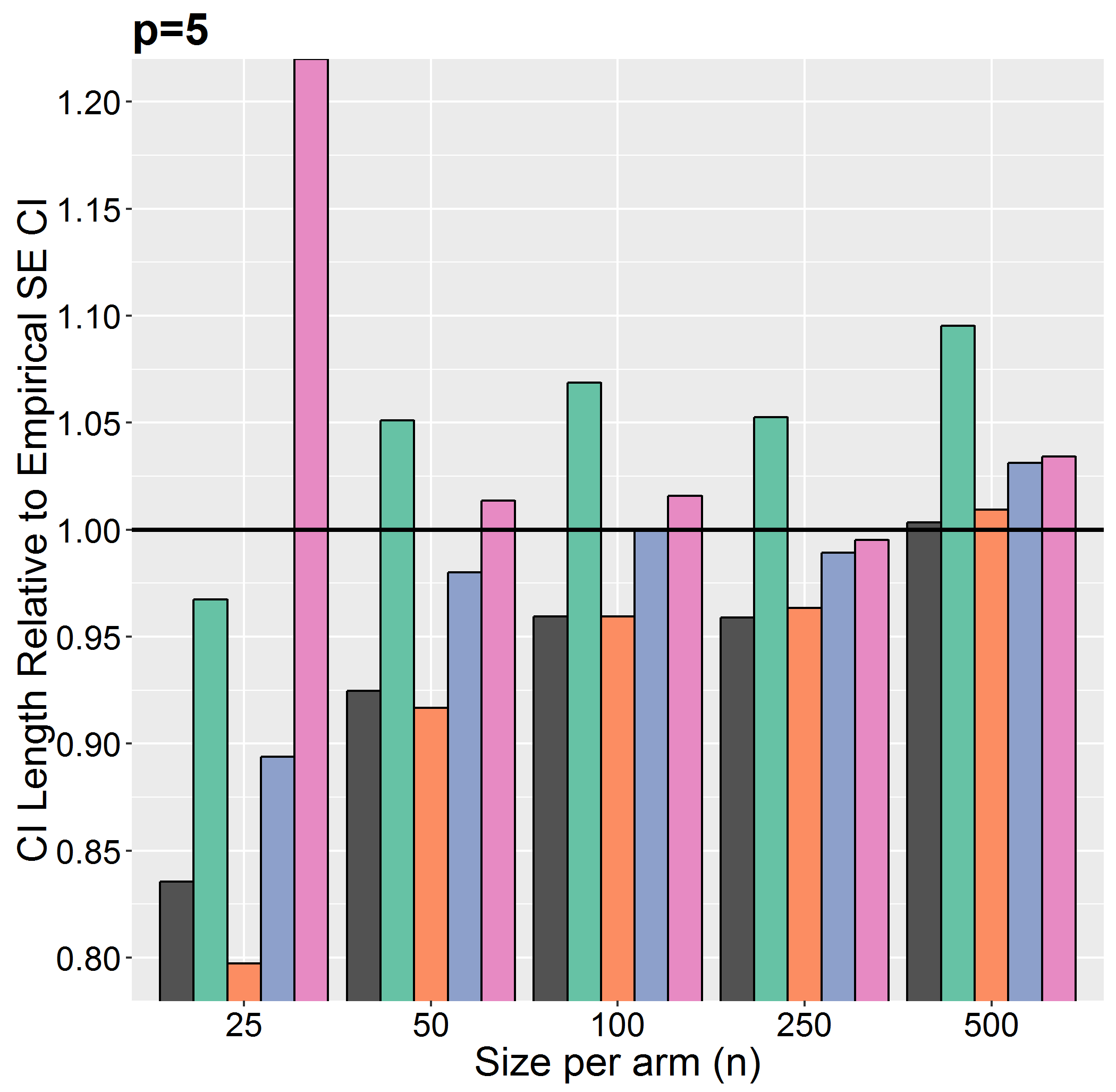} 
\includegraphics[scale=.29]{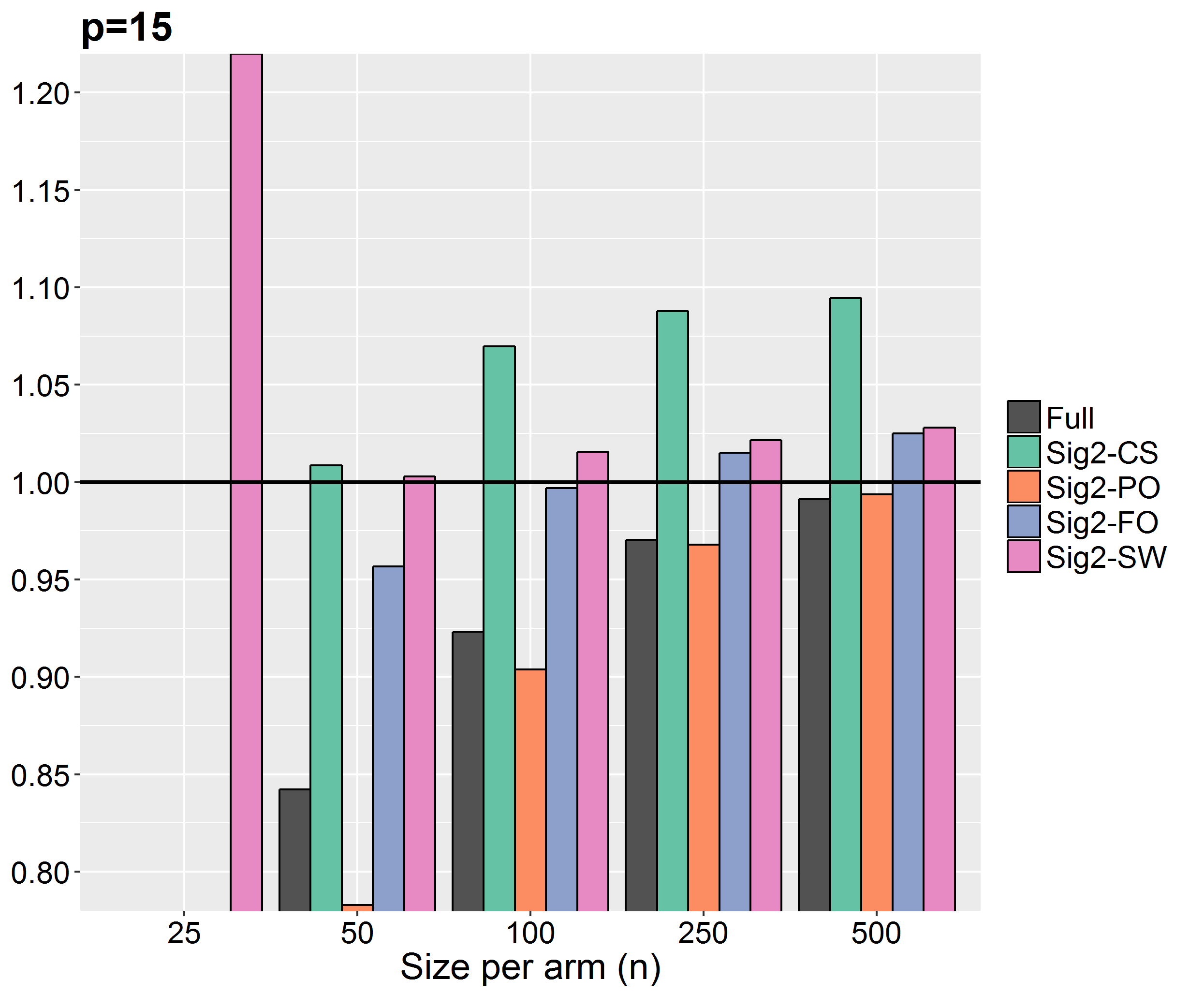}

Full: Based on full influence function, Sig2-CS: Based on $\sigmahat^2_{g,cs}$, Sig2-PO: Based on $\sigmahat^2_{g,po}$, Sig2-FO: Based on $\sigmahat^2_{g,fo}$, Sig2-SW: Based on $\sigmahat^2_{g,sw}$.
\end{center}
\end{figure}

The bias results are reported in Table \ref{t:biaslogit}.  Similar patterns of bias occur as when estimating $\Delta$.  BUC and STC incur substantial bias in scenarios with confounding, whereas MAIC has negligible bias except with extremely low sample size case with $n=25$ per arm where no method is reliable.  The CI performance results are presented in Figure \ref{f:covlogit}.  CIs based on $\sigmahat^2_{g,fo}$ and $\sigmahat^2_{g,sw}$ achieve near nominal coverage when $n\geq 50$ per arm and their lengths do not surpass those based on the empirical estimate by more than 5\% for most $n$.  CIs based on $\sigmahat^2_{g,po}$ exhibit good coverage when sample size is large relative to $p$ but tends to undercover when $n$ is relatively small.  CIs based on $\sigmahat^2_{g,cs}$ are the most conservative.

\begin{table}[htbp]
  \centering
  \caption{Percent bias of estimators of $\Delta_g$ by degree of confounding, total size ($N$), and number of covariates ($p$).}
    \scalebox{0.85}{
    \begin{tabular}{rrccc|ccc|ccc}
    \toprule
          &       & \multicolumn{3}{c}{No Confounding} & \multicolumn{3}{c}{Moderate Confounding} & \multicolumn{3}{c}{Severe Confounding} \\
    \hline
    \multicolumn{1}{c}{\textbf{Size}} & \multicolumn{1}{c}{\textbf{Estimator}} & \textbf{p=5} & \textbf{p=10} & \textbf{p=15} & \textbf{p=5} & \textbf{p=10} & \textbf{p=15} & \textbf{p=5} & \textbf{p=10} & \textbf{p=15} \\ \hline
    \multicolumn{1}{c}{\multirow{4}[2]{*}{$n=25$}} & \multicolumn{1}{c}{BUC} & 5\%   & 3\%   & 2\%   & 36\%  & 32\%  & 41\%  & 50\%  & 45\%  & 53\% \\
    \multicolumn{1}{c}{} & \multicolumn{1}{c}{STC} & 1315\% & 11064\% & 3752\% & 1816\% & 9903\% & 3392\% & 2498\% & 9066\% & 2951\% \\
    \multicolumn{1}{c}{} & MAIC-NAB & 14\%  & 27\%  & 365\% & 20\%  & 24\%  & 322\% & 23\%  & 41\%  & 362\% \\
    \multicolumn{1}{c}{} & MAIC-ACB & 0\%   & -7\%  & -122\% & 4\%   & -11\% & -44\% & -3\%  & 2\%   & -173\% \\ \hline
    \multicolumn{1}{c}{\multirow{4}[2]{*}{$n=50$}} & \multicolumn{1}{c}{BUC} & 3\%   & 2\%   & 0\%   & 28\%  & 32\%  & 33\%  & 44\%  & 42\%  & 47\% \\
    \multicolumn{1}{c}{} & \multicolumn{1}{c}{STC} & 31\%  & 146\% & 10724\% & 42\%  & 346\% & 11908\% & 53\%  & 702\% & 11294\% \\
    \multicolumn{1}{c}{} & MAIC-NAB & 4\%   & 8\%   & 6\%   & 7\%   & 9\%   & 15\%  & 7\%   & 10\%  & 10\% \\
    \multicolumn{1}{c}{} & MAIC-ACB & 3\%   & 0\%   & -4\%  & 0\%   & 4\%   & 7\%   & -1\%  & -2\%  & 0\% \\ \hline
    \multicolumn{1}{c}{\multirow{4}[2]{*}{$n=100$}} & \multicolumn{1}{c}{BUC} & 3\%   & 2\%   & -1\%  & 30\%  & 29\%  & 28\%  & 44\%  & 46\%  & 46\% \\
    \multicolumn{1}{c}{} & \multicolumn{1}{c}{STC} & 14\%  & 28\%  & 66\%  & 23\%  & 38\%  & 101\% & 33\%  & 45\%  & 209\% \\
    \multicolumn{1}{c}{} & MAIC-NAB & 3\%   & 3\%   & 2\%   & 3\%   & 4\%   & 4\%   & 6\%   & 3\%   & 6\% \\
    \multicolumn{1}{c}{} & MAIC-ACB & 3\%   & 1\%   & -1\%  & 2\%   & 2\%   & 1\%   & 1\%   & 2\%   & 2\% \\ \hline
    \multicolumn{1}{c}{\multirow{4}[2]{*}{$n=250$}} & \multicolumn{1}{c}{BUC} & 1\%   & 2\%   & 1\%   & 29\%  & 29\%  & 27\%  & 42\%  & 44\%  & 46\% \\
    \multicolumn{1}{c}{} & \multicolumn{1}{c}{STC} & 6\%   & 10\%  & 21\%  & 17\%  & 21\%  & 29\%  & 23\%  & 29\%  & 39\% \\
    \multicolumn{1}{c}{} & MAIC-NAB & 2\%   & 2\%   & 2\%   & 2\%   & 2\%   & 1\%   & 0\%   & 2\%   & 2\% \\
    \multicolumn{1}{c}{} & MAIC-ACB & 1\%   & 2\%   & 0\%   & 1\%   & 2\%   & 0\%   & -2\%  & 1\%   & 2\% \\ \hline
    \multicolumn{1}{c}{\multirow{4}[2]{*}{$n=500$}} & \multicolumn{1}{c}{BUC} & 0\%   & 0\%   & 1\%   & 28\%  & 27\%  & 28\%  & 41\%  & 43\%  & 44\% \\
    \multicolumn{1}{c}{} & \multicolumn{1}{c}{STC} & 3\%   & 5\%   & 9\%   & 14\%  & 16\%  & 21\%  & 21\%  & 25\%  & 29\% \\
    \multicolumn{1}{c}{} & MAIC-NAB & 1\%   & 1\%   & 1\%   & 1\%   & 0\%   & 1\%   & -1\%  & 1\%   & 2\% \\
    \multicolumn{1}{c}{} & MAIC-ACB & 0\%   & 0\%   & 1\%   & 0\%   & -1\%  & 1\%   & -2\%  & 0\%   & 1\% \\
    \bottomrule
    \end{tabular}%
  \label{t:biaslogit}%
}
BUC: Method of Bucher et al, STC: Simulated treatment comparison, MAIC-NAB: Non-anchor-based MAIC, MAIC-ACB: Anchor-based MAIC.
\end{table}%

\section{Alternative Sampling Schemes}
\label{apd:altsamp}
When estimating the standard error of $\Delthat$, a way to avoid the issue of the lack of full IPD is to consider $\bmustr_{\bX_2}$ and $\mustr_2$ as fixed parameters.
That is, one may consider $\bmustr_{\bX_2}= \bXbar_{2}$ or $\mustr_2 = \Ybar_{22}$.  Under either of these assumptions, the asymptotic variance $\sigma^2$ would exclude terms involving either $\varphitld_i^{\bmu_{\bX_2}}$ or $\varphi_i^{\mu_2}$, respectively.  Inference based on these assumptions clearly then would not acknowledge sampling variability from estimating these parameters, which may or may not be justified depending on the context of the problem.  In this article we primarily focus on the more difficult case where estimates from the AGD are considered to be random.

A related issue regarding sampling is that the allocation of patients among trials in practice may be constrained such that each trial and arm enrolls a fixed, or nearly fixed,  number of patients.  While formally this sampling scheme differs from the problem setup, which does not assume any sample size constraints, it can be accommodated as a simple extension that does not impact the asymptotic analysis. In particular, one could consider patients to be 
sub-sampled by trial and arm to meet size constraints after initially being sampled without constraint from the super-population.  Under the assumptions on the initial sample required when sampling without constraint, the proposed procedures for inference about $\Delta$ based on $\Delthat$ would still be valid using the sub-sampled data as long as the sub-sampling was random by trial and arm.

More concretely, suppose $\Nstr$ total patients were initially sampled without constraint. Let $R_i\in \{ 0,1\}$ for $i=1,\ldots,\Nstr$ be an indicator of whether each patient is sub-sampled into the final sample.  If a fixed $n_{tz}$ number of patients were enrolled into arm $z$ of trial $t$, the constraint on the sub-sampling is that $\sum_{i=1}^{\Nstr} R_i I(T_i=t,Z_i=z) = n_{tz}$, for $t=1,2$ and $z=0,1,2$.  If sub-sampling was random by trial and arm, then:
\begin{eq}
R \indep \{ \bX, Y(0), Y(1), Y(2)\} | T,Z.
\end{eq}
When Assumption \ref{a:randm} holds in the initial sample, then we also have that 
$R \indep \{ \bX, Y(0), Y(1), Y(2)\} | T$.  But under this independence:
\begin{eq*}
E\{Y(2)|T=2\} &= E\{Y(2)|T=2,R=1\}, \text{ and}\\
E\{Y(1)|T=2\} &= E\{Y(1)|T=2,R=1\}.
\end{eq*}
Moreover, Assumptions \ref{a:randm}-\ref{a:ignor} can be shown to still hold conditional on $R=1$, provided they hold in the initial sample.  The same arguments to identify $\Delta$ from Section \ref{ss:ident} thus hold under distributions that are conditional on $R=1$.  $\Delthat$ will then be consistent for $\Delta$ when the sub-sampled data is used, if $\omega(\bX)$ among the sub-sampled data can be identified and estimated.  But if trial assignment among the initial sample follows a logistic regression model as in \eqref{e:trlmod}, then:
\begin{eq*}
logitP(T=2|\bX,R=1) = \balph_T\trans\bXvec,
\end{eq*}
where $\balph_T = (\alpha_{0,T},\balph_1)$.  That is, the trial assignment among the sub-sample \emph{still} follows a logistic regression model with a possibly different intercept $\alpha_{0,T}$.  This is analogous to the result that the probability of an outcome given covariates among those sampled in a case-control study still follows the same logistic regression with a different intercept when the prospective model is logistic regression \citep{prentice1979logistic}.  The arguments for identification of $\omega(\bX)$ in Sections \ref{ss:ident} are thus also still valid, and estimation can proceed from solving \eqref{e:alphee} among the sub-sampled data.

\section{Proofs of Theorems}
\label{apd:proofs}

\subsection{Proof of Theorem \ref{thm:alphIF}}
Let the estimating equation from \eqref{e:alphee} be denoted:
\begin{eq}
\bU_N^{\balph_1}(\balph_1,\bmu_{\bX_2}) = N^{-1}\sum_{i=1}^N exp\{ \balph_1\trans(\bX_i - \bmu_{\bX_2})\}(\bX_i - \bmu_{\bX_2})I(T_i = 1).
\end{eq}
We will first outline the consistency of $\balphhat_1$ for $\balphstr_1$.   It can be shown that there exists a $C_N = O_p(1)$ such that:
\begin{eq}
&\sup_{\balph_1}\norm{\bU_N^{\balph_1}(\balph_1,\bmuhat_{\bX_2}) - \bU_N^{\balphtld_1,1}(\balph_1,\bmustr_{\bX_2})} \leq C_N \norm{\bmuhat_{\bX_2}-\bmustr_{\bX_2}} \\
&= o_p(1),
\end{eq}
by using that $\bU_N^{\balph_1}(\balph_1,\bmu_{\bX_2})$ is continuously differentiable in $\bmu_{\bX_2}$, the parameter space for $(\balph_1\trans,\bmu_{\bX_2}\trans)\trans$, denoted by $\Theta$, is compact, and $\bmuhat_{\bX_2}$ is a consistent estimator for $\bmustr_{\bX_2}$.  Let $\Theta_{\balph_1}$ be the parameter space for $\balph_1$. Moreover, it can be shown that there exists a $D_N = O_p(1)$ such that for any $\widetilde{\balphtld}_1,\balphtld_1 \in \Theta_{\balph_1}$:
\begin{eq}
\norm{\bU_N^{\balph_1}(\widetilde{\balphtld}_1,\bmustr_{\bX_2}) - \bU_N^{\balph_1}(\balphtld_1,\bmustr_{\bX_2})} \leq D_N \norm{\widetilde{\balphtld}_1 -\balphtld_1},
\end{eq}
using that $\bU_N^{\balph_1}(\balph_1,\bmustr_{\bX_2})$ is continuously differentiable in $\balph_1$ and $\Theta_{\balph_1}$ is compact.  Let $\bU^{\balph_1}(\balph_1,\bmu_{\bX_2}) = E[ exp\{ \balph_1\trans(\bX_i - \bmu_{\bX_2})\}(\bX_i-\bmu_{\bX_2})I(T_i=1)]$.  Since $\bU_N^{\balph_1}(\balph_1,\bmustr_{\bX_2}) \overset{p}{\to}\bU^{\balph_1}(\balph_1,\bmustr_{\bX_2})$ pointwise for each $\balph_1 \in \Theta_{\balph_1}$, we have:
\begin{eq}
\sup_{\balph_1}\norm{\bU_N^{\balph_1}(\balph_1,\bmustr_{\bX_2}) - \bU^{\balph_1}(\balph_1,\bmustr_{\bX_2})} = o_p(1)
\end{eq}
{ using Lemma 2.9 of \cite{newey1994large}.}  This verifies that:
\begin{eq}
&\sup_{\balph_1}\norm{\bU_N^{\balph_1}(\balph_1,\bmuhat_{\bX_2}) - \bU^{\balph_1}(\balph_1,\bmustr_{\bX_2})} \\
&\leq \sup_{\balph_1}\norm{\bU_N^{\balph_1}(\balph_1,\bmuhat_{\bX_2}) - \bU_N^{\balph_1}(\balph_1,\bmustr_{\bX_2})} + \sup_{\balph_1}\norm{\bU_N^{\balph_1}(\balph_1,\bmustr_{\bX_2}) - \bU^{\balph_1}(\balph_1,\bmustr_{\bX_2})} \\
&= o_p(1).
\end{eq}
Now since $ E[ exp\{ \balph_1\trans(\bX_i - \bmu_{\bX_2})\}I(T_i=1)]$ is strictly convex in $\balph_1$, $\bU^{\balph_1}(\balph_1,\bmu_{\bX_2})$ has a unique solution in $\balphtld_1$ for a given $\bmu_{\bX_2}$.  Hence $\balphhat_1 \overset{p}{\to}\balphstr_1$ by Theorem 5.9 of \cite{van2000asymptotic}.
We now turn to obtaining the influence function for $\balphhat_1$.  Let $U_{N,j}^{\balph_1}(\balph_1,\bmu_{\bX_2})$ denote the $j$-th component of $\bU_N^{\balph_1}(\balph_1,\bmu_{\bX_2})$, for $j=1,\ldots,p$.  An expansion of $U_{N,j}^{\balph_1}(\balphhat_1,\bmuhat_{\bX_2})$ around $(\balphstr_1,\bmustr_{\bX_2})$ yields:
\begin{eq}
&U_{N,j}^{\balph_1}(\balphhat_1,\bmuhat_{\bX_2}) = U_{N,j}^{\balph_1}(\balphstr_1,\bmustr_{\bX_2}) + \ddbalphT U_{N,j}^{\balph_1}(\balphstr_1,\bmustr_{\bX_2}) (\balphhat_1 - \balphstr_1) \\
&\quad +\ddbmuT U_{N,j}^{\balph_1}(\balphstr_1,\bmustr_{\bX_2}) (\bmuhat_{\bX_2}- \bmustr_{\bX_2}) + \ddbalphh U_{N,j}^{\balph_1}(\balphtld_1,\bmustr_{\bX_2}) (\balphhat_1 - \balphstr_1)^{\otimes 2} \\
&\quad +\ddbmuu U_{N,j}^{\balph_1}(\balphhat_1,\bmutld_{\bX_2}) (\bmuhat_{\bX_2}- \bmustr_{\bX_2})^{\otimes 2} + \ddbalphmu U_{N,j}^{\balph_1}(\widetilde{\balphtld}_1,\bmustr_{\bX_2})(\balphhat_1-\balphstr_1)(\bmuhat_{\bX_2}-\bmustr_{\bX_2}),
\end{eq}
where $\balphtld_1$ and $\widetilde{\balphtld}_1$ are intermediates on the line segment between $\balphhat_1$ and $\balphstr_1$, and $\bmutld_{\bX_2}$ is an intermediate between $\bmuhat_{\bX_2}$ and $\bmustr_{\bX_2}$.  Using that $U_{N,j}^{\balph_1}(\balph_1,\bmu_{\bX_2})$ is continuously differentiable in $(\balph_1\trans,\bmu_{\bX_2}\trans)\trans$ and $\Theta$ is compact:
\begin{eq}
&\ddbalphh U_{N,j}^{\balph_1}(\balphtld_1,\bmustr_{\bX_2}) = O_p(1), \ddbmuu U_{N,j}^{\balph_1}(\balphhat_1,\bmutld_{\bX_2}) = O_p(1), \\
&\ddbalphmu U_{N,j}^{\balph_1}(\widetilde{\balphtld}_1,\bmustr_{\bX_2}) = O_p(1).
\end{eq}
Moreover, we have:
\begin{eq}
&\ddbalphT U_{N,j}^{\balph_1}(\balphstr_1,\bmustr_{\bX_2}) = E\left[exp\{ \balphstrtrans_1 (\bX-\bmustr_{\bX_2})\} (X_j - \mustr_{\bX_2,j})(\bX-\bmustr_{\bX_2})\trans I(T=1)\right] + o_p(1) \\
&\qquad = -J_{j}^{\balph_1}(\balphstr_1,\bmustr_{\bX_2}) + o_p(1) \\
&\ddbmuT U_{N,j}^{\balph_1}(\balphstr_1,\bmustr_{\bX_2}) = -E\left[ exp\{ \balphstrtrans_1(\bX-\bmustr_{\bX_2})\}I(T=1)\bone_j\trans\right] + o_p(1),
\end{eq}
where $X_j$ and $\mu_{\bX_2,j}$ denote the $j$-th element of $\bX$ and $\bmu_{\bX_2}$ and $\bone_j$ denotes a $p\times 1$ vector that is $1$ in the $j$-th position and $0$ in the other positions.  We also used that $\balphstr_1$ is the solution to $\bU^{\balph_1}(\balph_1,\bmustr_{\bX_2})=\bzero$.

Now for each $j=1,\ldots,p$, since $U_{N,j}^{\balph_1}(\balphhat_1,\bmuhat_{\bX_2}) = 0$, re-arranging from above yields:
\begin{eq}
&\left\{-J_j^{\balph_1}(\balphstr_1,\bmustr_{\bX_2}) + o_p(1)\right\} N^{1/2}(\balphhat_1 - \balphstr_1)  = - N^{1/2}U_{N,j}^{\balph_1}(\balphstr_1,\bmustr_{\bX_2}) \\
&\quad -\ddbmuT U_{N,j}^{\balph_1}(\balphstr_1,\bmustr_{\bX_2}) N^{1/2}(\bmuhat_{\bX_2}- \bmustr_{\bX_2}) + o_p(1),
\end{eq}
where we use that $\muhat_{\bX_2}$ is $N^{1/2}$-consistent.  Considering the $p$ components simultaneously yields:
\begin{eq}
&\left\{ -\bJ^{\balph_1}(\balphstr_1,\bmustr_{\bX_2}) + o_p(1)\right\} N^{1/2}(\balphhat_1-\balphstr_1) = -N^{1/2}\bU_{N}^{\balph_1}(\balphstr_1,\bmustr_{\bX_2})  \\
&\quad  + E\left[ exp\{ \balphstrtrans_1(\bX-\bmustr_{\bX_2})\}I(T=1)\bone_{p\times p}\trans\right]N^{1/2} (\bmuhat_{\bX_2}-\bmustr_{\bX_2}) + o_p(1).
\end{eq}
Since $\bmuhat_{\bX_2} = \left\{ \sum_{i=1}^N \bX_i I(T_i=2)\right\}/\left\{ \sum_{i=1}^N I(T_i=2)\right\}$, its influence function is given by:
\begin{eq}
N^{1/2}(\bmuhat_{\bX_2}- \bmustr_{\bX_2}) = N^{-1/2}\sum_{i=1}^N (\bX_i - \bmustr_{\bX_2})\frac{I(T_i = 2)}{P(T_i = 2)} + o_p(1).
\end{eq}
Finally, since $\bJ^{\balph_1}(\balphstr_1,\bmustr_{\bX_2})$ is non-singular:
\begin{eq}
&N^{1/2}(\balphhat_1-\balphstr_1) = \bJ^{\balph_1}(\balphstr_1,\bmustr_{\bX_2})^{-1} \Big( N^{1/2}\bU_{N}^{\balph_1}(\balphstr_1,\bmustr_{\bX_2}) \\
&\quad - E\left[ exp\{ \balphstrtrans_1(\bX-\bmustr_{\bX_2})\}I(T=1)\right]N^{-1/2}\sum_{i=1}^N (\bX_i - \bmustr_{\bX_2})\frac{I(T_i = 2)}{P(T_i = 2)} \Big)+ o_p(1) \\
&\quad = N^{-1/2}\sum_{i=1}^N \bJ^{\balph_1}(\balphstr_1,\bmustr_{\bX_2})^{-1} \left\{ \bU_i^{\balph_1}(\balphstr_1,\bmustr_{\bX_2}) + \bU_i^{\bmu_{\bX_2}}(\balphstr_1,\bmustr_{\bX_2})\right\} + o_p(1)\\
&\quad = N^{-1/2} \sum_{i=1}^N \bvarphi_i^{\balph_1}(\balphstr_1,\bmustr_{\bX_2}) + o_p(1).
\end{eq}

\subsection{Proof of Theorem \ref{thm:DeltIF}}
Let the estimating equation associated with $\Delthat$ be:
\begin{eq}
U^{\Delta}_N(\Delta,\mu_1) = N^{-1}\sum_{i=1}^N (\mu_1- Y_i - \Delta) I(T_i = 2,Z_i=2).
\end{eq}

We will directly identify the influence function expansion for $\Delthat$, which implies that $\Delthat \overset{p}{\to}\Deltstr$, when the identification assumptions hold and the trial assignment is correctly specified.  First we decompose $U_N^{\Delta}(\Delthat,\muhat_1)$:
\begin{eq}
U_N^{\Delta}(\Delthat,\muhat_1) &= U_N^{\Delta}(\Deltstr,\mustr_1) + U_N^{\Delta}(\Delthat,\mustr_1) - U_N^{\Delta}(\Deltstr,\mustr_1) + U_N^{\Delta}(\Delthat,\muhat_1)- U_N^{\Delta}(\Delthat,\mustr_1) \\
&= U_N^{\Delta}(\Deltstr,\mustr_1) + N^{-1}\sum_{i=1}^N - I(T_i = 2,Z_i =2) (\Delthat - \Deltstr) \\
&\qquad + N^{-1}\sum_{i=1}^N I(T_i =2, Z_i = 2) (\muhat_1 - \mustr_1).
\end{eq}

Since $U_N^{\Delta}(\Delthat,\muhat_1) = 0$, re-arranging from above:
\begin{eq}
&N^{1/2}(\Delthat - \Deltstr) = N^{1/2}\left\{ U_N^{\Delta}(\Deltstr,\mustr_1) + N^{-1} \sum_{i=1}^N I(T_i = 2, Z_i = 2)(\muhat_1 - \mustr_1)\right\} \\
&\qquad\qquad\qquad\qquad  \Big/ \left\{ N^{-1}\sum_{i=1}^N I(T_i = 2, Z_i =2)\right\} \\
&\qquad = N^{-1/2}\sum_{i=1}^N (\mustr_1 - Y_i -\Deltstr) \frac{I(Z_i = 2,T_i = 2)}{P(Z_i=2,T_i=2)} + N^{1/2}(\muhat_1 - \mu_1) + o_p(1).
\end{eq}

Now let the estimating equation for $\muhat_1$ be denoted:
\begin{eq}
U_N^{\mu_1}(\mu_1,\balph_1) = N^{-1}\sum_{i=1}^N (Y_i - \mu_1)exp( \balph_1\trans\bX_i)I(T_i=1,Z_i=1).
\end{eq}

We can decompose $U_N^{\mu_1}(\muhat_1,\balphhat_1)$ as:
\begin{eq}
&U_N^{\mu_1}(\muhat_1,\balphhat_1) = U_N^{\mu_1}(\mustr_1,\balphstr_1) + U_N^{\mu_1}(\muhat_1,\balphstr_1) - U_N^{\mu_1}(\mustr_1,\balphstr_1) + U_N^{\mu_1}(\muhat_1,\balphhat_1) - U_N^{\mu_1}(\muhat_1,\balphstr_1)\\
&= U_N^{\mu_1}(\mustr_1,\balphstr_1) - N^{-1}\sum_{i=1}^N exp(\balphstrtrans_1 \bX_i) I(T_i = 1,Z_i = 1) (\muhat_1 - \mustr_1) \\
&\qquad + \ddbalphT U_N^{\mu_1}(\mustr_1,\balphstr_1)(\balphhat_1 - \balphstr_1)+ \ddbalphh U_N^{\mu_1}(\muhat_1,\balphtld_1)(\balphhat_1 - \balphstr_1)^{\otimes 2} \\
&\qquad\qquad + \ddmu1\ddbalphT U_N^{\mu_1}(\mutld_1,\balphstr_1)(\balphhat_1-\balphstr_1)(\muhat_1 - \mu_1),
\end{eq}
where $\balphtld_1$ and $\mutld_1$ are such that $\norm{\balphtld_1 - \balphstr_1} \leq \norm{\balphhat_1 - \balphstr_1}$ and $\norm{\mutld_1 - \mustr_1} \leq \norm{\muhat_1 - \mustr_1}$.  Now using that $U_N^{\mu_1}(\mu_1,\balph_1)$ is twice continuously differentaible and $\Theta$ is compact, it can be shown that:
\begin{eq}
&\ddbalphh U_N^{\mu_1}(\muhat_1,\balphtld_1) = O_p(1) \text{ and }\ddmu1\ddbalphT U_N^{\mu_1}(\mutld_1,\balphstr_1) = O_p(1).
\end{eq}
Since $U_N^{\mu_1}(\muhat_1,\balphhat_1)=0$, re-arranging yields:
\begin{eq} \label{e:DelthatIFintermed}
&\left\{ N^{-1}\sum_{i=1}^N exp(\balphstrtrans_1\bX_i)I(T_i=1,Z_i=1) + o_p(1)\right\}N^{1/2}(\muhat_1 - \mustr_1) \\
&\qquad= N^{1/2} \left\{ U_N^{\mu_1}(\mustr_1,\balphstr_1) + \ddbalphT U_N^{\mu_1}(\mustr_1,\balphstr_1)(\balphhat_1 - \balphstr_1)\right\} + o_p(1).
\end{eq}

Now substituting the influence function expansion from Theorem \ref{thm:alphIF} yields that:
\begin{eq}
N^{1/2}(\muhat_1 - \mustr_1) &= N^{-1/2}\sum_{i=1}^N J^{\mu_1}(\balphstr_1)^{-1}(Y_i - \mustr_1)exp(\balphstrtrans_1\bX_i)I(T_i=1,Z_i=1) \\
&\qquad + J^{\mu_1}(\balphstr_1)^{-1}\bCtld_1(\mustr_1,\balphstr_1)\trans\bvarphi_i^{\balph_1}(\balphstr_1,\bmustr_{\bX_2}) + o_p(1).
\end{eq}

Returning to \eqref{e:DelthatIFintermed}, we conclude that:
\begin{eq}
N^{1/2}(\Delthat-\Deltstr) &= N^{-1/2}\sum_{i=1}^N \varphi_i^{\mu_2}(\Deltstr,\mustr_1) + \varphi_i^{\mu_1}(\mustr_1,\balphstr_1) \\
&\qquad + J^{\mu_1}(\balphstr_1)^{-1}\bCtld_1(\mustr_1,\balphstr_1)\trans\bvarphi_i^{\balph_1}(\balphstr_1,\bmustr_{\bX_2}) + o_p(1) \\
&= N^{-1/2}\sum_{i=1}^N \varphi_i(\Deltstr,\mustr_1,\balphstr_1,\bmustr_{\bX_2}) + o_p(1).
\end{eq}

\subsection{Proof of Lemma \ref{lem:varsmp}}
From direct calculation:
\begin{eq}
&Var\{ \varphitld_i^{\balph_1}(\balph_1,\mu_1,\bmu_{\bX_2})\} = J^{\mu_1}(\balph_1)^{-2} \bCtld_1(\mu_1,\balph_1) \bJ^{\balph_1}(\balph_1,\bmu_{\bX_2})^{-1} \\
&\qquad Var\left\{ \bU_i^{\balph_1}(\balph_1,\bmu_{\bX_2})\right\} \bJ^{\balph_1}(\balph_1,\bmu_{\bX_2})^{-1}  \bCtld_1(\mu_1,\balph_1) \\
&= J^{\mu_1}(\balph_1)^{-2} \bCtld_1(\mu_1,\balph_1)\trans \bJ^{\balph_1}(\balph_1,\bmu_{\bX_2})^{-1} \\ 
&\qquad E\{ (\bX_i-\bmu_{\bX_2})(\bX_i-\bmu_{\bX_2})\trans e^{2\balph_1\trans(\bX_i-\bmu_{\bX_2})} I(T_i=1)\} \bJ^{\balph_1}(\balph_1,\bmu_{\bX_2})^{-1}  \bCtld_1(\mu_1,\balph_1)\\
&Cov\{\varphi_i^{\mu_1}(\mu_1,\balph_1) ,\varphitld_i^{\balph_1}(\balph_1,\mu_1,\bmu_{\bX_2})\}  = J^{\mu_1}(\balph_1)^{-2} \bCtld_1(\mu_1,\balph_1)\trans  \bJ^{\balph_1}(\balph_1,\bmu_{\bX_2})^{-1} \\
&\qquad E\{ \bU_i^{\balph_1}(\balph_1,\bmu_{\bX_2}) (Y_i - \mu_1)e^{\balph_1\trans\bX_i}I(Z_i = 1,T_i=1)\}\\
&= J^{\mu_1}(\balph_1)^{-2} \bCtld_1(\mu_1,\balph_1)\trans  \bJ^{\balph_1}(\balph_1,\bmu_{\bX_2})^{-1} e^{\balph_1\trans\bmu_{\bX_2}}\\ 
&\qquad E\{ (\bX_i - \bmu_{\bX_2})(Y_i-\mu_1)e^{2\balph_1\trans(\bX_i - \bmu_{\bX_2})} I(Z_i=1,T_i=1)\}\\
&Var\{ \varphitld_i^{\bmu_{\bX_2}}(\bmu_{\bX_2},\mu_1,\balph_1)\} = J^{\mu_1}(\balph_1)^{-2} \bCtld_1(\mu_1,\balph_1) \bJ^{\balph_1}(\balph_1,\bmu_{\bX_2})^{-1} \\
&\qquad Var\left\{ \bU_i^{\bmu_{\bX_2}}(\balph_1,\bmu_{\bX_2})\right\} \bJ^{\balph_1}(\balph_1,\bmu_{\bX_2})^{-1}  \bCtld_1(\mu_1,\balph_1) \\
&= \frac{E\left [ e^{ \balph_1\trans(\bX_i-\bmu_{\bX_2})} I(T_i=1)\right]^2}{J^{\mu_1}(\balph_1)^{2} P(T_i=2)} \bCtld_1(\mu_1,\balph_1)\trans \bJ^{\balph_1}(\balph_1,\bmu_{\bX_2})^{-1} \\ 
&\qquad Var( \bX_i|T_i=2) \bJ^{\balph_1}(\balph_1,\bmu_{\bX_2})^{-1}  \bCtld_1(\mu_1,\balph_1)\\
&Cov\{ \varphi^{\mu_2}(\Delta,\mu_1),\varphitld_i^{\bmu_{\bX_2}}(\balph_1,\mu_1,\bmu_{\bX_2})\} = \frac{E\left [ e^{ \balph_1\trans(\bX_i-\bmu_{\bX_2})} I(T_i=1)\right]}{J^{\mu_1}(\balph_1)P(T_i = 2)} \\
&\qquad \bCtld_1(\mu_1,\balph_1)\trans \bJ^{\balph_1}(\balph_1,\bmu_{\bX_2})^{-1}\bCtld_2.
\end{eq}
In the case where \eqref{a:randm}, \eqref{a:consi}, \eqref{a:ignor}, and \eqref{a:posit} hold and \eqref{e:trlmod} is correctly specified, note that:
\begin{eq}
\bCtld_1(\mu_1,\balph_1) &= E\left\{ \bX_i (Y_i - \mu_1)\omega(\bX_i)e^{-\alpha_0} I(Z_i=1,T_i=1)\right\} \\
&= E\left\{ \bX_i (Y_i(1) - \mu_1)P(T_i=2|\bX_i)\right\} e^{-\alpha_0}P(Z_i=1|T_i=1)\\
&= \bC_1 e^{-\alpha_0}P(Z_i=1|T_i=1)P(T_i=2)\\
\bJ^{\balph_1}(\balph_1,\bmu_{\bX_2}) &= -E\left[ (\bX_i - \bmu_{\bX_2})(\bX_i - \bmu_{\bX_2})\trans \omega(\bX_i)e^{-\alpha_0-\balph_1\trans\bmu_{\bX_2}}I(T_i=1)\right] \\
&= -E\left[ (\bX_i - \bmu_{\bX_2})(\bX_i - \bmu_{\bX_2})\trans P(T_i=2|\bX_i)\right]e^{-\alpha_0-\balph_1\trans\bmu_{\bX_2}} \\
&= -Var(\bX_i|T_i=2)P(T_i=2)e^{-\alpha_0-\balph_1\trans\bmu_{\bX_2}},
\end{eq}
and:
\begin{eq}
&E\left [ e^{ \balph_1\trans(\bX_i-\bmu_{\bX_2})} I(T_i=1)\right] = E\left [ \omega(\bX_i)e^{-\alpha_0-\balph_1\trans\bmu_{\bX_2}} I(T_i=1)\right]  \\
&\qquad\qquad= P(T_i=2)e^{-\alpha_0-\balph_1\trans\bmu_{\bX_2}}  \\
&J^{\mu_1}(\balph_1) = E\{ \omega(\bX_i)I(Z_i=1,T_i=1)\}e^{-\alpha_0} \\
&\qquad\qquad =P(T_i=2)P(Z_i=1|T_i=1)e^{-\alpha_0}.
\end{eq}
Consequently under these assumptions after evaluating parameters at the truth and simplification:
\begin{eq}
&Var( \varphitld_i^{\balph_1}) = P(T_i=2)^{-1} \bC_1\trans Var(\bX_i|T_i=2)^{-1} \\ 
&\qquad E\{ (\bX_i-\bmustr_{\bX_2})(\bX_i-\bmustr_{\bX_2})\trans \omega(\bX_i) | T_i=2\} Var(\bX_i|T_i=2)^{-1} \bC_1\\
&Cov(\varphi_i^{\mu_1} ,\varphitld_i^{\balph_1})  = -P(T_i=2)^{-1}\bC_1Var(\bX_i|T_i=2)^{-1}\\ 
&\qquad E\{ (\bX_i - \bmustr_{\bX_2})(Y_i(1)-\mustr_1)\omega(\bX_i) |T_i=2\} \\
&Var( \varphitld_i^{\bmu_{\bX_2}}) 
= P(T_i=2)^{-1}\bC_1\trans Var(\bX_i|T_i=2)^{-1}\bC_1 \\
&Cov( \varphi_i^{\mu_2},\varphitld_i^{\bmu_{\bX_2}}) 
= -P(T_i=2)^{-1}\bC_1\trans Var(\bX_i|T_i=2)^{-1}\bC_2.
\end{eq}
\section*{Acknowledgement}
The authors would like to thank the patients and families participating in the DMD-PRO-01 and tadalafil DMD trials, 
as well as our collaborators at the Collaborative Trajectory Analysis Project (cTAP), CureDuchenne, and Eli Lilly, for 
making the data available for this research.

\bibliographystyle{imsart-nameyear}
\bibliography{ref}

\begin{thebibliography}{53}

\bibitem[\protect\citeauthoryear{Adjei, Christian and
  Ivy}{2009}]{adjei2009novel}
\begin{barticle}[author]
\bauthor{\bsnm{Adjei},~\bfnm{Alex~A}\binits{A.~A.}},
  \bauthor{\bsnm{Christian},~\bfnm{Michaele}\binits{M.}} \AND
  \bauthor{\bsnm{Ivy},~\bfnm{Percy}\binits{P.}}
(\byear{2009}).
\btitle{Novel designs and end points for phase II clinical trials}.
\bjournal{Clinical Cancer Research}
\bvolume{15}
\bpages{1866--1872}.
\end{barticle}
\endbibitem

\bibitem[\protect\citeauthoryear{Altman et~al.}{2005}]{altman2005indirect}
\begin{barticle}[author]
\bauthor{\bsnm{Altman},~\bfnm{D}\binits{D.}},
  \bauthor{\bsnm{Song},~\bfnm{F}\binits{F.}},
  \bauthor{\bsnm{Sakarovitch},~\bfnm{C}\binits{C.}},
  \bauthor{\bsnm{Deeks},~\bfnm{J}\binits{J.}},
  \bauthor{\bsnm{D'Amico},~\bfnm{R}\binits{R.}},
  \bauthor{\bsnm{Bradburn},~\bfnm{M}\binits{M.}} \AND
  \bauthor{\bsnm{Eastwood},~\bfnm{A~Glenny~A}\binits{A.~G.~A.}}
(\byear{2005}).
\btitle{Indirect comparisons of competing interventions.}
\bjournal{Health Technology Assessment}.
\end{barticle}
\endbibitem

\bibitem[\protect\citeauthoryear{Bland and Altman}{2000}]{bland2000odds}
\begin{barticle}[author]
\bauthor{\bsnm{Bland},~\bfnm{J~Martin}\binits{J.~M.}} \AND
  \bauthor{\bsnm{Altman},~\bfnm{Douglas~G}\binits{D.~G.}}
(\byear{2000}).
\btitle{The odds ratio}.
\bjournal{Bmj}
\bvolume{320}
\bpages{1468}.
\end{barticle}
\endbibitem

\bibitem[\protect\citeauthoryear{Brookhart
  et~al.}{2006}]{brookhart2006variable}
\begin{barticle}[author]
\bauthor{\bsnm{Brookhart},~\bfnm{M~Alan}\binits{M.~A.}},
  \bauthor{\bsnm{Schneeweiss},~\bfnm{Sebastian}\binits{S.}},
  \bauthor{\bsnm{Rothman},~\bfnm{Kenneth~J}\binits{K.~J.}},
  \bauthor{\bsnm{Glynn},~\bfnm{Robert~J}\binits{R.~J.}},
  \bauthor{\bsnm{Avorn},~\bfnm{Jerry}\binits{J.}} \AND
  \bauthor{\bsnm{St{\"u}rmer},~\bfnm{Til}\binits{T.}}
(\byear{2006}).
\btitle{Variable selection for propensity score models}.
\bjournal{American journal of epidemiology}
\bvolume{163}
\bpages{1149--1156}.
\end{barticle}
\endbibitem

\bibitem[\protect\citeauthoryear{Bucher et~al.}{1997}]{bucher1997results}
\begin{barticle}[author]
\bauthor{\bsnm{Bucher},~\bfnm{Heiner~C}\binits{H.~C.}},
  \bauthor{\bsnm{Guyatt},~\bfnm{Gordon~H}\binits{G.~H.}},
  \bauthor{\bsnm{Griffith},~\bfnm{Lauren~E}\binits{L.~E.}} \AND
  \bauthor{\bsnm{Walter},~\bfnm{Stephen~D}\binits{S.~D.}}
(\byear{1997}).
\btitle{The results of direct and indirect treatment comparisons in
  meta-analysis of randomized controlled trials}.
\bjournal{Journal of clinical epidemiology}
\bvolume{50}
\bpages{683--691}.
\end{barticle}
\endbibitem

\bibitem[\protect\citeauthoryear{Caro and Ishak}{2010}]{caro2010no}
\begin{barticle}[author]
\bauthor{\bsnm{Caro},~\bfnm{J~Jaime}\binits{J.~J.}} \AND
  \bauthor{\bsnm{Ishak},~\bfnm{K~Jack}\binits{K.~J.}}
(\byear{2010}).
\btitle{No head-to-head trial? Simulate the missing arms}.
\bjournal{Pharmacoeconomics}
\bvolume{28}
\bpages{957--967}.
\end{barticle}
\endbibitem

\bibitem[\protect\citeauthoryear{{EMA CHMP}}{2006}]{committee2006guideline}
\begin{barticle}[author]
\bauthor{\bsnm{{EMA CHMP}}}
(\byear{2006}).
\btitle{Guideline on clinical trials in small populations}.
\bjournal{London: EMEA}.
\end{barticle}
\endbibitem

\bibitem[\protect\citeauthoryear{{EMA CHMP}}{2018}]{ema2018kymriah}
\begin{bmisc}[author]
\bauthor{\bsnm{{EMA CHMP}}}
(\byear{2018}).
\btitle{Assessment Report: Kymriah (International non-proprietary name:
  tisagenlecleucel)}.
\bhowpublished{\url{https://www.ema.europa.eu/en/documents/assessment-report/kymriah-epar-public-assessment-report_en.pdf}}.
\bnote{[Online; accessed 1-April-2020]}.
\end{bmisc}
\endbibitem

\bibitem[\protect\citeauthoryear{Cohen}{2013}]{cohen2013statistical}
\begin{bbook}[author]
\bauthor{\bsnm{Cohen},~\bfnm{Jacob}\binits{J.}}
(\byear{2013}).
\btitle{Statistical power analysis for the behavioral sciences}.
\bpublisher{Academic press}.
\end{bbook}
\endbibitem

\bibitem[\protect\citeauthoryear{De~Luna, Waernbaum and
  Richardson}{2011}]{de2011covariate}
\begin{barticle}[author]
\bauthor{\bsnm{De~Luna},~\bfnm{Xavier}\binits{X.}},
  \bauthor{\bsnm{Waernbaum},~\bfnm{Ingeborg}\binits{I.}} \AND
  \bauthor{\bsnm{Richardson},~\bfnm{Thomas~S}\binits{T.~S.}}
(\byear{2011}).
\btitle{Covariate selection for the nonparametric estimation of an average
  treatment effect}.
\bjournal{Biometrika}
\bvolume{98}
\bpages{861--875}.
\end{barticle}
\endbibitem

\bibitem[\protect\citeauthoryear{Emery, Muntoni and
  Quinlivan}{2015}]{emery2015duchenne}
\begin{bbook}[author]
\bauthor{\bsnm{Emery},~\bfnm{Alan~EH}\binits{A.~E.}},
  \bauthor{\bsnm{Muntoni},~\bfnm{Francesco}\binits{F.}} \AND
  \bauthor{\bsnm{Quinlivan},~\bfnm{Rosaline~CM}\binits{R.~C.}}
(\byear{2015}).
\btitle{Duchenne muscular dystrophy}.
\bpublisher{OUP Oxford}.
\end{bbook}
\endbibitem

\bibitem[\protect\citeauthoryear{Fay and Graubard}{2001}]{fay2001small}
\begin{barticle}[author]
\bauthor{\bsnm{Fay},~\bfnm{Michael~P}\binits{M.~P.}} \AND
  \bauthor{\bsnm{Graubard},~\bfnm{Barry~I}\binits{B.~I.}}
(\byear{2001}).
\btitle{Small-sample adjustments for Wald-type tests using sandwich
  estimators}.
\bjournal{Biometrics}
\bvolume{57}
\bpages{1198--1206}.
\end{barticle}
\endbibitem

\bibitem[\protect\citeauthoryear{{FDA}}{2001}]{us2001guidance}
\begin{barticle}[author]
\bauthor{\bsnm{{FDA}}}
(\byear{2001}).
\btitle{Guidance for Industry. E 10 choice of control group and related issues
  in clinical trials}.
\bjournal{US Department of Health and Human Services, Federal Drug
  Administration.}
\end{barticle}
\endbibitem

\bibitem[\protect\citeauthoryear{FDA}{2006}]{us2006fda}
\begin{bmisc}[author]
\bauthor{\bsnm{FDA}}
(\byear{2006}).
\btitle{Drug Approval Package: Myozyme (Alglucosidase Alfa)}.
\bhowpublished{\url{https://www.accessdata.fda.gov/drugsatfda_docs/nda/2006/125141s000_MyozymeTOC.cfm
  }}.
\bnote{[Online; accessed 11-June-2019]}.
\end{bmisc}
\endbibitem

\bibitem[\protect\citeauthoryear{FDA}{2017}]{us2017fda}
\begin{bmisc}[author]
\bauthor{\bsnm{FDA}}
(\byear{2017}).
\btitle{FDA Approves First Treatment for a Form of Batten Disease}.
\bhowpublished{\url{https://www.fda.gov/newsevents/newsroom/pressannouncements/ucm555613.htm}}.
\bnote{[Online; accessed 11-June-2019]}.
\end{bmisc}
\endbibitem

\bibitem[\protect\citeauthoryear{Goemans
  et~al.}{2016}]{goemans2016individualized}
\begin{barticle}[author]
\bauthor{\bsnm{Goemans},~\bfnm{Nathalie}\binits{N.}},
  \bauthor{\bsnm{Vanden~Hauwe},~\bfnm{Marleen}\binits{M.}},
  \bauthor{\bsnm{Signorovitch},~\bfnm{James}\binits{J.}},
  \bauthor{\bsnm{Swallow},~\bfnm{Elyse}\binits{E.}},
  \bauthor{\bsnm{Song},~\bfnm{Jinlin}\binits{J.}} \AND
  \bauthor{\bsnm{Project},~\bfnm{Collaborative
  Trajectory~Analysis}\binits{C.~T.~A.}}
(\byear{2016}).
\btitle{Individualized prediction of changes in 6-minute walk distance for
  patients with Duchenne muscular dystrophy}.
\bjournal{PloS one}
\bvolume{11}.
\end{barticle}
\endbibitem

\bibitem[\protect\citeauthoryear{Hahn}{1998}]{hahn1998role}
\begin{barticle}[author]
\bauthor{\bsnm{Hahn},~\bfnm{Jinyong}\binits{J.}}
(\byear{1998}).
\btitle{On the role of the propensity score in efficient semiparametric
  estimation of average treatment effects}.
\bjournal{Econometrica}
\bpages{315--331}.
\end{barticle}
\endbibitem

\bibitem[\protect\citeauthoryear{Hainmueller}{2012}]{hainmueller2012entropy}
\begin{barticle}[author]
\bauthor{\bsnm{Hainmueller},~\bfnm{Jens}\binits{J.}}
(\byear{2012}).
\btitle{Entropy balancing for causal effects: A multivariate reweighting method
  to produce balanced samples in observational studies}.
\bjournal{Political Analysis}
\bvolume{20}
\bpages{25--46}.
\end{barticle}
\endbibitem

\bibitem[\protect\citeauthoryear{Hartman et~al.}{2015}]{hartman2015sample}
\begin{barticle}[author]
\bauthor{\bsnm{Hartman},~\bfnm{Erin}\binits{E.}},
  \bauthor{\bsnm{Grieve},~\bfnm{Richard}\binits{R.}},
  \bauthor{\bsnm{Ramsahai},~\bfnm{Roland}\binits{R.}} \AND
  \bauthor{\bsnm{Sekhon},~\bfnm{Jasjeet~S}\binits{J.~S.}}
(\byear{2015}).
\btitle{From sample average treatment effect to population average treatment
  effect on the treated: combining experimental with observational studies to
  estimate population treatment effects}.
\bjournal{Journal of the Royal Statistical Society: Series A (Statistics in
  Society)}
\bvolume{178}
\bpages{757--778}.
\end{barticle}
\endbibitem

\bibitem[\protect\citeauthoryear{ICH}{2000}]{guideline2000choice}
\begin{barticle}[author]
\bauthor{\bsnm{ICH}}
(\byear{2000}).
\btitle{Choice of control group and related issues in clinical trials E10}.
\end{barticle}
\endbibitem

\bibitem[\protect\citeauthoryear{Imai and Ratkovic}{2014}]{imai2014covariate}
\begin{barticle}[author]
\bauthor{\bsnm{Imai},~\bfnm{Kosuke}\binits{K.}} \AND
  \bauthor{\bsnm{Ratkovic},~\bfnm{Marc}\binits{M.}}
(\byear{2014}).
\btitle{Covariate balancing propensity score}.
\bjournal{Journal of the Royal Statistical Society: Series B (Statistical
  Methodology)}
\bvolume{76}
\bpages{243--263}.
\end{barticle}
\endbibitem

\bibitem[\protect\citeauthoryear{Ishak, Proskorovsky and
  Benedict}{2015}]{ishak2015simulation}
\begin{barticle}[author]
\bauthor{\bsnm{Ishak},~\bfnm{K~Jack}\binits{K.~J.}},
  \bauthor{\bsnm{Proskorovsky},~\bfnm{Irina}\binits{I.}} \AND
  \bauthor{\bsnm{Benedict},~\bfnm{Agnes}\binits{A.}}
(\byear{2015}).
\btitle{Simulation and matching-based approaches for indirect comparison of
  treatments}.
\bjournal{Pharmacoeconomics}
\bvolume{33}
\bpages{537--549}.
\end{barticle}
\endbibitem

\bibitem[\protect\citeauthoryear{Jansen et~al.}{2011}]{jansen2011interpreting}
\begin{barticle}[author]
\bauthor{\bsnm{Jansen},~\bfnm{Jeroen~P}\binits{J.~P.}},
  \bauthor{\bsnm{Fleurence},~\bfnm{Rachael}\binits{R.}},
  \bauthor{\bsnm{Devine},~\bfnm{Beth}\binits{B.}},
  \bauthor{\bsnm{Itzler},~\bfnm{Robbin}\binits{R.}},
  \bauthor{\bsnm{Barrett},~\bfnm{Annabel}\binits{A.}},
  \bauthor{\bsnm{Hawkins},~\bfnm{Neil}\binits{N.}},
  \bauthor{\bsnm{Lee},~\bfnm{Karen}\binits{K.}},
  \bauthor{\bsnm{Boersma},~\bfnm{Cornelis}\binits{C.}},
  \bauthor{\bsnm{Annemans},~\bfnm{Lieven}\binits{L.}} \AND
  \bauthor{\bsnm{Cappelleri},~\bfnm{Joseph~C}\binits{J.~C.}}
(\byear{2011}).
\btitle{Interpreting indirect treatment comparisons and network meta-analysis
  for health-care decision making: report of the ISPOR Task Force on Indirect
  Treatment Comparisons Good Research Practices: part 1}.
\bjournal{Value in Health}
\bvolume{14}
\bpages{417--428}.
\end{barticle}
\endbibitem

\bibitem[\protect\citeauthoryear{Kang and Schafer}{2007}]{kang2007demystifying}
\begin{barticle}[author]
\bauthor{\bsnm{Kang},~\bfnm{Joseph~DY}\binits{J.~D.}} \AND
  \bauthor{\bsnm{Schafer},~\bfnm{Joseph~L}\binits{J.~L.}}
(\byear{2007}).
\btitle{Demystifying double robustness: A comparison of alternative strategies
  for estimating a population mean from incomplete data}.
\bjournal{Statistical science}
\bpages{523--539}.
\end{barticle}
\endbibitem

\bibitem[\protect\citeauthoryear{Kauermann and
  Carroll}{2001}]{kauermann2001note}
\begin{barticle}[author]
\bauthor{\bsnm{Kauermann},~\bfnm{G{\"o}ran}\binits{G.}} \AND
  \bauthor{\bsnm{Carroll},~\bfnm{Raymond~J}\binits{R.~J.}}
(\byear{2001}).
\btitle{A note on the efficiency of sandwich covariance matrix estimation}.
\bjournal{Journal of the American Statistical Association}
\bvolume{96}
\bpages{1387--1396}.
\end{barticle}
\endbibitem

\bibitem[\protect\citeauthoryear{Kong}{1992}]{kong1992note}
\begin{barticle}[author]
\bauthor{\bsnm{Kong},~\bfnm{Augustine}\binits{A.}}
(\byear{1992}).
\btitle{A note on importance sampling using standardized weights}.
\bjournal{University of Chicago, Dept. of Statistics, Tech. Rep}
\bvolume{348}.
\end{barticle}
\endbibitem

\bibitem[\protect\citeauthoryear{Li, Morgan and
  Zaslavsky}{2017}]{li2017balancing}
\begin{barticle}[author]
\bauthor{\bsnm{Li},~\bfnm{Fan}\binits{F.}},
  \bauthor{\bsnm{Morgan},~\bfnm{Kari~Lock}\binits{K.~L.}} \AND
  \bauthor{\bsnm{Zaslavsky},~\bfnm{Alan~M}\binits{A.~M.}}
(\byear{2017}).
\btitle{Balancing covariates via propensity score weighting}.
\bjournal{Journal of the American Statistical Association}
\bpages{1--11}.
\end{barticle}
\endbibitem

\bibitem[\protect\citeauthoryear{Lu and Ades}{2004}]{lu2004combination}
\begin{barticle}[author]
\bauthor{\bsnm{Lu},~\bfnm{Guobing}\binits{G.}} \AND
  \bauthor{\bsnm{Ades},~\bfnm{AE}\binits{A.}}
(\byear{2004}).
\btitle{Combination of direct and indirect evidence in mixed treatment
  comparisons}.
\bjournal{Statistics in medicine}
\bvolume{23}
\bpages{3105--3124}.
\end{barticle}
\endbibitem

\bibitem[\protect\citeauthoryear{Lunceford and
  Davidian}{2004}]{lunceford2004stratification}
\begin{barticle}[author]
\bauthor{\bsnm{Lunceford},~\bfnm{Jared~K}\binits{J.~K.}} \AND
  \bauthor{\bsnm{Davidian},~\bfnm{Marie}\binits{M.}}
(\byear{2004}).
\btitle{Stratification and weighting via the propensity score in estimation of
  causal treatment effects: a comparative study}.
\bjournal{Statistics in medicine}
\bvolume{23}
\bpages{2937--2960}.
\end{barticle}
\endbibitem

\bibitem[\protect\citeauthoryear{Mazzone et~al.}{2016}]{mazzone2016timed}
\begin{barticle}[author]
\bauthor{\bsnm{Mazzone},~\bfnm{Elena~S}\binits{E.~S.}},
  \bauthor{\bsnm{Coratti},~\bfnm{Giorgia}\binits{G.}},
  \bauthor{\bsnm{Sormani},~\bfnm{Maria~Pia}\binits{M.~P.}},
  \bauthor{\bsnm{Messina},~\bfnm{Sonia}\binits{S.}},
  \bauthor{\bsnm{Pane},~\bfnm{Marika}\binits{M.}},
  \bauthor{\bsnm{D'Amico},~\bfnm{Adele}\binits{A.}},
  \bauthor{\bsnm{Colia},~\bfnm{Giulia}\binits{G.}},
  \bauthor{\bsnm{Fanelli},~\bfnm{Lavinia}\binits{L.}},
  \bauthor{\bsnm{Berardinelli},~\bfnm{Angela}\binits{A.}},
  \bauthor{\bsnm{Gardani},~\bfnm{Alice}\binits{A.}} \betal{et~al.}
(\byear{2016}).
\btitle{Timed rise from floor as a predictor of disease progression in Duchenne
  muscular dystrophy: an observational study}.
\bjournal{PloS one}
\bvolume{11}.
\end{barticle}
\endbibitem

\bibitem[\protect\citeauthoryear{Newey and McFadden}{1994}]{newey1994large}
\begin{barticle}[author]
\bauthor{\bsnm{Newey},~\bfnm{Whitney~K}\binits{W.~K.}} \AND
  \bauthor{\bsnm{McFadden},~\bfnm{Daniel}\binits{D.}}
(\byear{1994}).
\btitle{Large sample estimation and hypothesis testing}.
\bjournal{Handbook of econometrics}
\bvolume{4}
\bpages{2111--2245}.
\end{barticle}
\endbibitem

\bibitem[\protect\citeauthoryear{Nie et~al.}{2013}]{nie2013likelihood}
\begin{barticle}[author]
\bauthor{\bsnm{Nie},~\bfnm{Lei}\binits{L.}},
  \bauthor{\bsnm{Zhang},~\bfnm{Zhiwei}\binits{Z.}},
  \bauthor{\bsnm{Rubin},~\bfnm{Daniel}\binits{D.}},
  \bauthor{\bsnm{Chu},~\bfnm{Jianxiong}\binits{J.}} \betal{et~al.}
(\byear{2013}).
\btitle{Likelihood reweighting methods to reduce potential bias in
  noninferiority trials which rely on historical data to make inference}.
\bjournal{The Annals of Applied Statistics}
\bvolume{7}
\bpages{1796--1813}.
\end{barticle}
\endbibitem

\bibitem[\protect\citeauthoryear{Nixon, Bansback and
  Brennan}{2007}]{nixon2007using}
\begin{barticle}[author]
\bauthor{\bsnm{Nixon},~\bfnm{RM}\binits{R.}},
  \bauthor{\bsnm{Bansback},~\bfnm{Nick}\binits{N.}} \AND
  \bauthor{\bsnm{Brennan},~\bfnm{Alan}\binits{A.}}
(\byear{2007}).
\btitle{Using mixed treatment comparisons and meta-regression to perform
  indirect comparisons to estimate the efficacy of biologic treatments in
  rheumatoid arthritis}.
\bjournal{Statistics in medicine}
\bvolume{26}
\bpages{1237--1254}.
\end{barticle}
\endbibitem

\bibitem[\protect\citeauthoryear{Phillippo et~al.}{2016}]{phillippo20162016}
\begin{barticle}[author]
\bauthor{\bsnm{Phillippo},~\bfnm{D}\binits{D.}},
  \bauthor{\bsnm{Ades},~\bfnm{T}\binits{T.}},
  \bauthor{\bsnm{Dias},~\bfnm{S}\binits{S.}},
  \bauthor{\bsnm{Palmer},~\bfnm{S}\binits{S.}},
  \bauthor{\bsnm{Abrams},~\bfnm{KR}\binits{K.}} \AND
  \bauthor{\bsnm{Welton},~\bfnm{N}\binits{N.}}
(\byear{2016}).
\btitle{NICE DSU Technical Support Document 18: Methods for population-adjusted
  indirect comparisons in submissions to NICE. (Technical Support Documents).
  Decision Support Unit, ScHARR, University of Sheffield: NICE Decision Support
  Unit.}
\end{barticle}
\endbibitem

\bibitem[\protect\citeauthoryear{Pocock}{1976}]{pocock1976combination}
\begin{barticle}[author]
\bauthor{\bsnm{Pocock},~\bfnm{Stuart~J}\binits{S.~J.}}
(\byear{1976}).
\btitle{The combination of randomized and historical controls in clinical
  trials}.
\bjournal{Journal of chronic diseases}
\bvolume{29}
\bpages{175--188}.
\end{barticle}
\endbibitem

\bibitem[\protect\citeauthoryear{Prentice and
  Pyke}{1979}]{prentice1979logistic}
\begin{barticle}[author]
\bauthor{\bsnm{Prentice},~\bfnm{Ross~L}\binits{R.~L.}} \AND
  \bauthor{\bsnm{Pyke},~\bfnm{Ronald}\binits{R.}}
(\byear{1979}).
\btitle{Logistic disease incidence models and case-control studies}.
\bjournal{Biometrika}
\bvolume{66}
\bpages{403--411}.
\end{barticle}
\endbibitem

\bibitem[\protect\citeauthoryear{Ricotti et~al.}{2016}]{ricotti2016northstar}
\begin{barticle}[author]
\bauthor{\bsnm{Ricotti},~\bfnm{Valeria}\binits{V.}},
  \bauthor{\bsnm{Ridout},~\bfnm{Deborah~A}\binits{D.~A.}},
  \bauthor{\bsnm{Pane},~\bfnm{Marika}\binits{M.}},
  \bauthor{\bsnm{Main},~\bfnm{Marion}\binits{M.}},
  \bauthor{\bsnm{Mayhew},~\bfnm{Anna}\binits{A.}},
  \bauthor{\bsnm{Mercuri},~\bfnm{Eugenio}\binits{E.}},
  \bauthor{\bsnm{Manzur},~\bfnm{Adnan~Y}\binits{A.~Y.}} \AND
  \bauthor{\bsnm{Muntoni},~\bfnm{Francesco}\binits{F.}}
(\byear{2016}).
\btitle{The NorthStar Ambulatory Assessment in Duchenne muscular dystrophy:
  considerations for the design of clinical trials}.
\bjournal{J Neurol Neurosurg Psychiatry}
\bvolume{87}
\bpages{149--155}.
\end{barticle}
\endbibitem

\bibitem[\protect\citeauthoryear{Rotnitzky, Li and
  Li}{2010}]{rotnitzky2010note}
\begin{barticle}[author]
\bauthor{\bsnm{Rotnitzky},~\bfnm{Andrea}\binits{A.}},
  \bauthor{\bsnm{Li},~\bfnm{Lingling}\binits{L.}} \AND
  \bauthor{\bsnm{Li},~\bfnm{Xiaochun}\binits{X.}}
(\byear{2010}).
\btitle{A note on overadjustment in inverse probability weighted estimation}.
\bjournal{Biometrika}
\bvolume{97}
\bpages{997--1001}.
\end{barticle}
\endbibitem

\bibitem[\protect\citeauthoryear{Rubin and Thomas}{1996}]{rubin1996matching}
\begin{barticle}[author]
\bauthor{\bsnm{Rubin},~\bfnm{Donald~B}\binits{D.~B.}} \AND
  \bauthor{\bsnm{Thomas},~\bfnm{Neal}\binits{N.}}
(\byear{1996}).
\btitle{Matching using estimated propensity scores: relating theory to
  practice}.
\bjournal{Biometrics}
\bpages{249--264}.
\end{barticle}
\endbibitem

\bibitem[\protect\citeauthoryear{Signorovitch
  et~al.}{2010}]{signorovitch2010comparative}
\begin{barticle}[author]
\bauthor{\bsnm{Signorovitch},~\bfnm{James~E}\binits{J.~E.}},
  \bauthor{\bsnm{Wu},~\bfnm{Eric~Q}\binits{E.~Q.}},
  \bauthor{\bsnm{Andrew},~\bfnm{P~Yu}\binits{P.~Y.}},
  \bauthor{\bsnm{Gerrits},~\bfnm{Charles~M}\binits{C.~M.}},
  \bauthor{\bsnm{Kantor},~\bfnm{Evan}\binits{E.}},
  \bauthor{\bsnm{Bao},~\bfnm{Yanjun}\binits{Y.}},
  \bauthor{\bsnm{Gupta},~\bfnm{Shiraz~R}\binits{S.~R.}} \AND
  \bauthor{\bsnm{Mulani},~\bfnm{Parvez~M}\binits{P.~M.}}
(\byear{2010}).
\btitle{Comparative effectiveness without head-to-head trials}.
\bjournal{Pharmacoeconomics}
\bvolume{28}
\bpages{935--945}.
\end{barticle}
\endbibitem

\bibitem[\protect\citeauthoryear{Signorovitch
  et~al.}{2012}]{signorovitch2012matching}
\begin{barticle}[author]
\bauthor{\bsnm{Signorovitch},~\bfnm{James~E}\binits{J.~E.}},
  \bauthor{\bsnm{Sikirica},~\bfnm{Vanja}\binits{V.}},
  \bauthor{\bsnm{Erder},~\bfnm{M~Haim}\binits{M.~H.}},
  \bauthor{\bsnm{Xie},~\bfnm{Jipan}\binits{J.}},
  \bauthor{\bsnm{Lu},~\bfnm{Mei}\binits{M.}},
  \bauthor{\bsnm{Hodgkins},~\bfnm{Paul~S}\binits{P.~S.}},
  \bauthor{\bsnm{Betts},~\bfnm{Keith~A}\binits{K.~A.}} \AND
  \bauthor{\bsnm{Wu},~\bfnm{Eric~Q}\binits{E.~Q.}}
(\byear{2012}).
\btitle{Matching-adjusted indirect comparisons: a new tool for timely
  comparative effectiveness research}.
\bjournal{Value in Health}
\bvolume{15}
\bpages{940--947}.
\end{barticle}
\endbibitem

\bibitem[\protect\citeauthoryear{Signorovitch
  et~al.}{2013}]{signorovitch2013everolimus}
\begin{barticle}[author]
\bauthor{\bsnm{Signorovitch},~\bfnm{James}\binits{J.}},
  \bauthor{\bsnm{Swallow},~\bfnm{Elyse}\binits{E.}},
  \bauthor{\bsnm{Kantor},~\bfnm{Evan}\binits{E.}},
  \bauthor{\bsnm{Wang},~\bfnm{Xufang}\binits{X.}},
  \bauthor{\bsnm{Klimovsky},~\bfnm{Judith}\binits{J.}},
  \bauthor{\bsnm{Haas},~\bfnm{Tomas}\binits{T.}},
  \bauthor{\bsnm{Devine},~\bfnm{Beth}\binits{B.}} \AND
  \bauthor{\bsnm{Metrakos},~\bfnm{Peter}\binits{P.}}
(\byear{2013}).
\btitle{Everolimus and sunitinib for advanced pancreatic neuroendocrine tumors:
  a matching-adjusted indirect comparison}.
\bjournal{Experimental hematology \& oncology}
\bvolume{2}
\bpages{32}.
\end{barticle}
\endbibitem

\bibitem[\protect\citeauthoryear{Snapinn and Jiang}{2011}]{snapinn2011indirect}
\begin{barticle}[author]
\bauthor{\bsnm{Snapinn},~\bfnm{Steven}\binits{S.}} \AND
  \bauthor{\bsnm{Jiang},~\bfnm{Qi}\binits{Q.}}
(\byear{2011}).
\btitle{Indirect comparisons in the comparative efficacy and non-inferiority
  settings}.
\bjournal{Pharmaceutical statistics}
\bvolume{10}
\bpages{420--426}.
\end{barticle}
\endbibitem

\bibitem[\protect\citeauthoryear{Stuart}{2010}]{stuart2010matching}
\begin{barticle}[author]
\bauthor{\bsnm{Stuart},~\bfnm{Elizabeth~A}\binits{E.~A.}}
(\byear{2010}).
\btitle{Matching methods for causal inference: A review and a look forward}.
\bjournal{Statistical science: a review journal of the Institute of
  Mathematical Statistics}
\bvolume{25}
\bpages{1}.
\end{barticle}
\endbibitem

\bibitem[\protect\citeauthoryear{Stuart et~al.}{2011}]{stuart2011use}
\begin{barticle}[author]
\bauthor{\bsnm{Stuart},~\bfnm{Elizabeth~A}\binits{E.~A.}},
  \bauthor{\bsnm{Cole},~\bfnm{Stephen~R}\binits{S.~R.}},
  \bauthor{\bsnm{Bradshaw},~\bfnm{Catherine~P}\binits{C.~P.}} \AND
  \bauthor{\bsnm{Leaf},~\bfnm{Philip~J}\binits{P.~J.}}
(\byear{2011}).
\btitle{The use of propensity scores to assess the generalizability of results
  from randomized trials}.
\bjournal{Journal of the Royal Statistical Society: Series A (Statistics in
  Society)}
\bvolume{174}
\bpages{369--386}.
\end{barticle}
\endbibitem

\bibitem[\protect\citeauthoryear{Sutton et~al.}{2008}]{sutton2008use}
\begin{barticle}[author]
\bauthor{\bsnm{Sutton},~\bfnm{Alex}\binits{A.}},
  \bauthor{\bsnm{Ades},~\bfnm{AE}\binits{A.}},
  \bauthor{\bsnm{Cooper},~\bfnm{Nicola}\binits{N.}} \AND
  \bauthor{\bsnm{Abrams},~\bfnm{Keith}\binits{K.}}
(\byear{2008}).
\btitle{Use of indirect and mixed treatment comparisons for technology
  assessment}.
\bjournal{Pharmacoeconomics}
\bvolume{26}
\bpages{753--767}.
\end{barticle}
\endbibitem

\bibitem[\protect\citeauthoryear{Swallow et~al.}{2016}]{swallow2016daclatasvir}
\begin{barticle}[author]
\bauthor{\bsnm{Swallow},~\bfnm{Elyse}\binits{E.}},
  \bauthor{\bsnm{Song},~\bfnm{Jinlin}\binits{J.}},
  \bauthor{\bsnm{Yuan},~\bfnm{Yong}\binits{Y.}},
  \bauthor{\bsnm{Kalsekar},~\bfnm{Anupama}\binits{A.}},
  \bauthor{\bsnm{Kelley},~\bfnm{Caroline}\binits{C.}},
  \bauthor{\bsnm{Peeples},~\bfnm{Miranda}\binits{M.}},
  \bauthor{\bsnm{Mu},~\bfnm{Fan}\binits{F.}},
  \bauthor{\bsnm{Ackerman},~\bfnm{Peter}\binits{P.}} \AND
  \bauthor{\bsnm{Signorovitch},~\bfnm{James}\binits{J.}}
(\byear{2016}).
\btitle{Daclatasvir and sofosbuvir versus sofosbuvir and ribavirin in patients
  with chronic hepatitis C coinfected with HIV: a matching-adjusted indirect
  comparison}.
\bjournal{Clinical therapeutics}
\bvolume{38}
\bpages{404--412}.
\end{barticle}
\endbibitem

\bibitem[\protect\citeauthoryear{Van~der Vaart}{2000}]{van2000asymptotic}
\begin{bbook}[author]
\bauthor{\bparticle{Van~der} \bsnm{Vaart},~\bfnm{Aad~W}\binits{A.~W.}}
(\byear{2000}).
\btitle{Asymptotic statistics}
\bvolume{3}.
\bpublisher{Cambridge university press}.
\end{bbook}
\endbibitem

\bibitem[\protect\citeauthoryear{Wells et~al.}{2009}]{wells2009indirect}
\begin{barticle}[author]
\bauthor{\bsnm{Wells},~\bfnm{GA}\binits{G.}},
  \bauthor{\bsnm{Sultan},~\bfnm{SA}\binits{S.}},
  \bauthor{\bsnm{Chen},~\bfnm{L}\binits{L.}},
  \bauthor{\bsnm{Khan},~\bfnm{M}\binits{M.}} \AND
  \bauthor{\bsnm{Coyle},~\bfnm{D}\binits{D.}}
(\byear{2009}).
\btitle{Indirect evidence: indirect treatment comparisons in meta-analysis}.
\bjournal{Ottawa: Canadian Agency for Drugs and Technologies in Health}
\bpages{1--94}.
\end{barticle}
\endbibitem

\bibitem[\protect\citeauthoryear{Zeileis}{2004}]{zeileis2004econometric}
\begin{barticle}[author]
\bauthor{\bsnm{Zeileis},~\bfnm{Achim}\binits{A.}}
(\byear{2004}).
\btitle{Econometric computing with HC and HAC covariance matrix estimators}.
\end{barticle}
\endbibitem

\bibitem[\protect\citeauthoryear{Zhang et~al.}{2015}]{zhang2015new}
\begin{barticle}[author]
\bauthor{\bsnm{Zhang},~\bfnm{Zhiwei}\binits{Z.}},
  \bauthor{\bsnm{Nie},~\bfnm{Lei}\binits{L.}},
  \bauthor{\bsnm{Soon},~\bfnm{Guoxing}\binits{G.}} \AND
  \bauthor{\bsnm{Hu},~\bfnm{Zonghui}\binits{Z.}}
(\byear{2015}).
\btitle{New methods for treatment effect calibration, with applications to
  non-inferiority trials}.
\bjournal{Biometrics}.
\end{barticle}
\endbibitem

\bibitem[\protect\citeauthoryear{Zhao and Percival}{2017}]{zhao2017entropy}
\begin{barticle}[author]
\bauthor{\bsnm{Zhao},~\bfnm{Qingyuan}\binits{Q.}} \AND
  \bauthor{\bsnm{Percival},~\bfnm{Daniel}\binits{D.}}
(\byear{2017}).
\btitle{Entropy balancing is doubly robust}.
\bjournal{Journal of Causal Inference}
\bvolume{5}.
\end{barticle}
\endbibitem

\bibitem[\protect\citeauthoryear{Zubizarreta}{2015}]{zubizarreta2015stable}
\begin{barticle}[author]
\bauthor{\bsnm{Zubizarreta},~\bfnm{Jos{\'e}~R}\binits{J.~R.}}
(\byear{2015}).
\btitle{Stable weights that balance covariates for estimation with incomplete
  outcome data}.
\bjournal{Journal of the American Statistical Association}
\bvolume{110}
\bpages{910--922}.
\end{barticle}
\endbibitem

\end{thebibliography}

\end{document}